\documentclass[epsf,draftcopy,graphics,subfigure]{mn2e}
\usepackage{epsf,graphics,graphicx}
\usepackage[T1]{fontenc}
\usepackage{ae,aecompl}

\newif\ifAMStwofonts

\def\h{$^{\rm h}$}
\def\m{$^{\rm m}$}

\ifCUPmtlplainloaded \else \ifAMStwofonts \else 
  \def\upi{\pi} \def\umu{\mu} \def\upartial{\partial} \fi \fi
  \title[High resolution radio studies of HDF/HFF sources] {High
  Resolution Studies of Radio Sources in the Hubble Deep and Flanking
  Fields} 
\author[T. W. B. Muxlow et al.]  {T.W.B. Muxlow$^1$,
   A.M.S. Richards$^1$, S.T. Garrington$^1$, P.N. Wilkinson$^1$,
  \newauthor B. Anderson$^1$, E.A. Richards$^{2}$, D.J. Axon$^3$,
  E.B. Fomalont$^4$, K.I. Kellermann$^4$, \newauthor R.B. Partridge$^5$ and R.A. Windhorst$^6$\\
  $^1$MERLIN/VLBI National Facility, Jodrell Bank Observatory,
  University of Manchester, Macclesfield, Cheshire, SK11 9DL, UK.\\
  $^2$Department of Physics, Talladega College, Talledega, Alabama 35160, USA \\ 
  $^3$Department of Physics, Rochester Institute of Technology, 84 Lomb Memorial Drive, Rochester, NY 14623, USA.\\
  $^4$NRAO Edgemont Road, Charlottesville, Virginia 22903, USA.\\
  $^5$Department of Astronomy, Haverford College, Haverford, PA 19041, USA.\\
  $^6$Department of Physics and Astronomy, Arizona State University, Box 871504, Tempe, AZ 85287-1504, USA.\\
}
\pubyear{2004}
\begin{document}
\maketitle
\label{firstpage}
\begin{abstract}
Eighteen days of MERLIN data and 42 hours of A-array VLA data at 1.4
GHz have been combined to image a 10-arcmin field centred on the
Hubble Deep Field (HDF). This area also includes the Hubble Flanking
Fields (HFF). A complete sample of 92 radio sources with $S_{1.4}>40 $
$\mu$Jy was detected using the VLA data alone and then imaged with the
MERLIN+VLA combination. The combined images offer: i) higher angular
resolution (synthesised beams of diameter 0.2--0.5~arcsec); ii)
improved astrometric accuracy and iii) improved sensitivity compared
with VLA-only data. The images are amongst the most sensitive yet made
at 1.4 GHz, with rms noise levels of 3.3~$\mu$Jy beam$^{-1}$ in the
0.2-arcsec images. Virtually all the sources are resolved, with
angular sizes in the range 0.2 to 3~arcsec. The central 3-arcmin
square was imaged separately to search for sources down to
27~$\mu$Jy. No additional sources were detected, indicating that
sources fainter than 40 $\mu$Jy are heavily resolved with MERLIN and must
have typical angular sizes $>$0.5 arcsec. Radio sources associated
with compact galaxies have been used to align the HDF, the HFF and a
larger CFHT optical field, to the radio-based International Celestial
Reference Frame. The {\em HST} optical fields have been registered to
$<50$ mas in the HDF itself, and to $\leq$150 mas in the outer parts
of the HFF. We find a {\it statistical} association of very faint
($\ge 2$~$\mu$Jy) radio sources with optically bright HDF galaxies down
to $\sim 23$ mag.  Of the 92 radio sources above 40~$\mu$Jy,
$\sim$85~per cent are identified with galaxies brighter than {\em
I}~=~25 mag; the remaining 15~per cent are associated with optically
faint systems close to or beyond the HFF (or even the HDF) limit.  The
high astrometric accuracy and the ability of radio waves to penetrate
obscuring dust has led to the correct identification of several very
red, optically faint systems
including the the strongest sub-mm source in the HDF, HDF850.1.  On
the basis of their radio structures and spectra 72~per cent (66
sources) can be classified as starburst or AGN-type systems; the
remainder are unclassified. The
proportion of starburst systems increases with decreasing flux
density; below 100 $\mu$Jy $>$70~per cent of the sources are
starburst-type systems associated with major disc galaxies in the
redshift range 0.3 -- 1.3.  {\em Chandra} detections are associated
with 55 of the 92 radio sources but their X-ray flux densities do not
appear to be correlated with the radio flux densities or
morphologies. The most recent sub-mm results on the HDF and HFF do not
provide any unambiguous identifications with these latest radio data,
except for HDF850.1, but suggest at least three strong candidates.
\end{abstract}
\section{INTRODUCTION}
\label{sect-intro}
Radio observations allow us to probe the nuclear regions of galaxies
which are obscured by gas and dust in other wavebands. In addition,
radio observations with high angular resolution can distinguish, on
morphological grounds, between emission that is driven by
star-formation and that which is driven by Active Galactic Nuclei
(AGN). While sub-mm wavelengths provide one means of detecting
star-formation in the early Universe, through redshifted infra-red
radiation from dusty star-forming regions, current sub-mm instruments
do not have sufficient angular resolution to avoid source blending in
deep sub-mm fields, or the astrometric capability to avoid confusion
in crowded optical/NIR fields. Sensitive, high angular-resolution
radio observations are currently the best way to uniquely identify the
optical counterparts to this population. Furthermore, because of the
negative {\em K}-correction, dusty galaxies observed at sub-mm
wavelengths preferentially lie at high redshifts whereas radio
observations are sensitive to a wide range of redshifts and to a mix
of starburst and AGN activity and hence provide complementary
information on the population of distant galaxies.

The {\em Hubble Space Telescope} ({\em HST}) has been used to image a
window on the distant universe to very high sensitivity. The central
Hubble Deep Field (HDF) and surrounding Hubble Flanking Fields (HFF)
are complete to
{\em R} magnitudes of 29 and 25 
respectively (Williams et al.\/ 1996). The HDF region was selected to
avoid known bright galaxies in any observable waveband.


We report here the results from our high-resolution, high-sensitivity
1.4-GHz MERLIN+VLA study of the $\mu$Jy sources found in a 10-arcmin
field enclosing both the HDF and HFF, together with a more sensitive
search for radio sources in a 3-arcmin square field enclosing the HDF
itself. These represent amongst the most sensitive 1.4-GHz
observations yet made and together with their high angular resolution
(0.2~arcsec) and astrometric positional accuracy (tens of mas) these
data allow, for the first time, detailed imaging and accurate
identification of $\mu$Jy radio sources.

This work represents a major extension to the original 1.4-GHz VLA
results reported in Richards et al.\/ (1999) and Richards (2000)
providing better astrometry, more sensitive maps, and higher angular
resolution. It also complements and adds to the work at 8.4 GHz
reported by Fomalont et al.\/ (2002) and Richards et al.\/
(1998).

In Sections~\ref{sect-obs} and~\ref{sect-images}, we summarize the VLA
and MERLIN observations and the procedures adopted to make the radio
images. In Section~\ref{sect-align} we describe in some detail the
process of alignment of the optical and radio fields in order to
achieve a registration accuracy of $<$50 mas (HDF) to $<$150 mas
(HFF/outer 10-arcmin field). In Section~\ref{sect-10arcmin} we discuss
the individual radio/optical objects in the 10-arcmin field. In
Section~\ref{sect-3arcmin} we outline the different procedure adopted
to image the 3-arcmin field and discuss the statistical association of
very weak radio sources with optical objects down to {\em R}~=~25
mag. In Section~\ref{sect-properties} we discuss the structural
properties of the $\mu$Jy radio sources, their optical identifications
and redshifts, their luminosities and inferred star-formation
rates. In Section~\ref{sect-other} we make comparisons with other
observations in radio, sub-mm, infra-red, and X-ray wavebands; whilst
in Section~\ref{sect-faint} we discuss the optically faint systems.
We look forward to the possibilities opened up by the next generation
of telescopes in Section~\ref{sect-future-radio-obs} and our overall
conclusions from this deep radio study are presented in
Section~\ref{sect-concl}.

\section{THE RADIO OBSERVATIONS AND DATA REDUCTION}
\label{sect-obs}		
\subsection{The VLA observations}
\label{sect-vla}
The detailed description of the VLA observations and calibration are
given in Richards (2000) and we merely summarize the work here for
completeness. We observed an area centred on the Hubble Deep Field
and located at R.A.\/~12\h~36\m~49\fs4 Dec.\/ 
+62\degr~12\arcmin~58\farcs00 (J2000) with the VLA in A-array for a
total of 50 hours at 1.4 GHz. In order to minimise chromatic
aberration, we observed in `pseudo-continuum mode' with $7\times3.125$ MHz
channels centred on intermediate frequencies 1365 MHz and 1435 MHz,
frequency windows previously established to be relatively free of
radio frequency interference.

After time-averaging the data to 13 seconds and significant tapering
in the spatial frequency plane, preliminary 10-arcsec resolution maps
were made covering the field out to the first side-lobe of the primary
beam -- about 0\degr.8 from the field centre. These were then
searched for bright, confusing sources with flux densities whose
side-lobes might contaminate the noise characteristics of the inner
portion of the field.

Each confusing source above 0.5 mJy was imaged and heavily {\sc clean}ed
using the full resolution data set. Because of changes in the primary
beam response across our 44-MHz bandpass it was necessary to
deconvolve each of the confusing sources in spectral-line mode
separately from each frequency channel. Furthermore, the confusing
sources were independently deconvolved for each hand of circular
polarization; this was to account for the beam-squint of the VLA
antennas which produces small but significant differences for each
hand of polarization for sources on the edge of the primary beam. The
{\sc clean} components were then Fourier transformed and subtracted from the
visibility data which were then used to image the inner few arcmin
of the field. Following this procedure, the rms noise was found to be
about 50~per cent higher than expected from receiver noise alone. In
particular, a few side-lobes from particularly strong confusing sources
($S>$10 mJy) located near the half-power point of the primary beam were
still apparent.

By examining the images made from 30 minute segments of data, we
isolated a few time intervals where the visibility data appeared to be
corrupted, possibly due to incomplete confusing source subtraction
associated with telescope pointing errors, or perhaps low level
interference. We deleted all data segments where the rms noise
exceeded the mean value by greater than 50~per cent. This amounted to
about 7 hours of data and, in all, about 42 hours of high quality data
were used to construct the final images. In the final VLA images we
achieved an rms noise of 7.5 $\mu$Jy beam$^{-1}$, compared with a
theoretical value closer to 5 $\mu$Jy beam$^{-1}$. 92 sources with
flux densities $>$40 $\mu$Jy (5.3$\sigma$) were reliably detected in
the 10-arcmin $\times$ 10-arcmin field.

\subsection{The MERLIN observations}
\label{sect-merlin}
We observed the same position with the Multi-Element Radio Linked
Interferometer Network (MERLIN) at 1.4 GHz in February 1996 and April
1997 for a total of 18.23 days. These observations included the
76-metre diameter Lovell Telescope which increases the array
sensitivity by a factor $\sim$2.4 compared with identical observations
excluding the Lovell Telescope. Observations with a single
intermediate frequency, were centred on 1420.0 MHz with
$32\times0.5$-MHz channels to allow imaging over a wide field 10
arcmin on a side (comparable with the primary beam of the Lovell
Telescope HPBW=12.4 arcmin). Data from the correlator were recorded
every 4 seconds. As with the VLA data, correlator limitations resulted
in only parallel hands of circular polarization being correlated, thus
no linear polarization information is available.

The interferometric phases were calibrated by observing the nearby 0.4
Jy phase calibration source J124129+622041 with a duty cycle of 8:2
minutes (source:calibrator). The visibility amplitudes were calibrated
against 3C286 and B1803+784 with assumed flux densities of 14.774 Jy
and 1.914 Jy respectively (Baars et al.\/ 1977). Bandpass
calibration was performed using B1803+784. Visibilities corrupted by
instrumental phase errors, telescope errors, or external interference
were flagged and discarded. Finally, the visibilities were re-weighted
to reflect the differing sensitivities of the various telescopes
involved in the MERLIN array. Due to the high angular resolution of
MERLIN, its lack of short interferometer spacings, and the effect of
the primary beam of the Lovell Telescope, no significant problems with
confusing sources outside the 10-arcmin field were
encountered. However, the four strongest sources within the 10-arcmin
 field were mapped and their {\sc clean} components subtracted from
the MERLIN data, prior to the MERLIN+VLA imaging of the remaining 88
weaker sources detected by the VLA. In addition, the `subtracted
data' set was used for the imaging of the 3-arcmin field. The
frequency channels at the extreme ends of the passband are subject to
a loss of sensitivity due to bandwidth limitations on the microwave
radio links. The lowest rms noise value was found when excluding the
lowest three and the highest channel, resulting in a final bandwidth
of 14 MHz. The rms noise achieved in naturally-weighted images from the
MERLIN data alone was 5.9 $\mu$Jy beam$^{-1}$ which is within 10 per cent of the
theoretical value.

\section{THE RADIO IMAGES}
\label{sect-images}
\subsection{Data and image combination}
\label{sect-comb}
Since the correlators used for the observations of the HDF and HFF
were optimally configured for the two separate arrays, the MERLIN and
VLA visibility datasets are fundamentally different in structure,
containing differing numbers of intermediate frequencies, frequency
channels, and channel bandwidths. For this reason, combining the
visibility datasets in the {\sc aips} software package proved wholly
impracticable, not least due to the size of the visibility datasets
which, in combination, proved far too large for the software package
as implemented to handle. The data were therefore combined in the
sky-plane rather than the {\it uv} plane.

For all the 92 radio sources detected by the VLA in the 10-arcmin
field, individual `postage-stamp' maps were made centred on each
source from both the VLA and MERLIN datasets separately. No
deconvolution was performed prior to any image combination; thus the
`dirty maps' and `dirty beams' were made from each
naturally-weighted dataset separately. In order to retain optimum
Fourier Transform gridding for each dataset, a different cell-size was
selected for each array (MERLIN 0.05~arcsec, VLA 0.4~arcsec). The VLA
`dirty maps' and `beams' were then re-gridded to the cell-size of
the MERLIN images and then averaged with them to produce combination
`dirty maps' and a combination `dirty beam'. We gave equal
statistical weight to each input image since the MERLIN and VLA images
were of comparable sensitivity. This process is equivalent to
combining the data sets prior to Fourier transformation into the sky
plane. The central quarters of these combination images were then
deconvolved with a conventional {\sc clean} algorithm (H\"{o}gbom, 1974)
resulting in deconvolved images 25.6~arcsec on a side.

\begin{figure*}
\rotatebox{-90}{
\resizebox{9cm}{!}{
\epsfbox{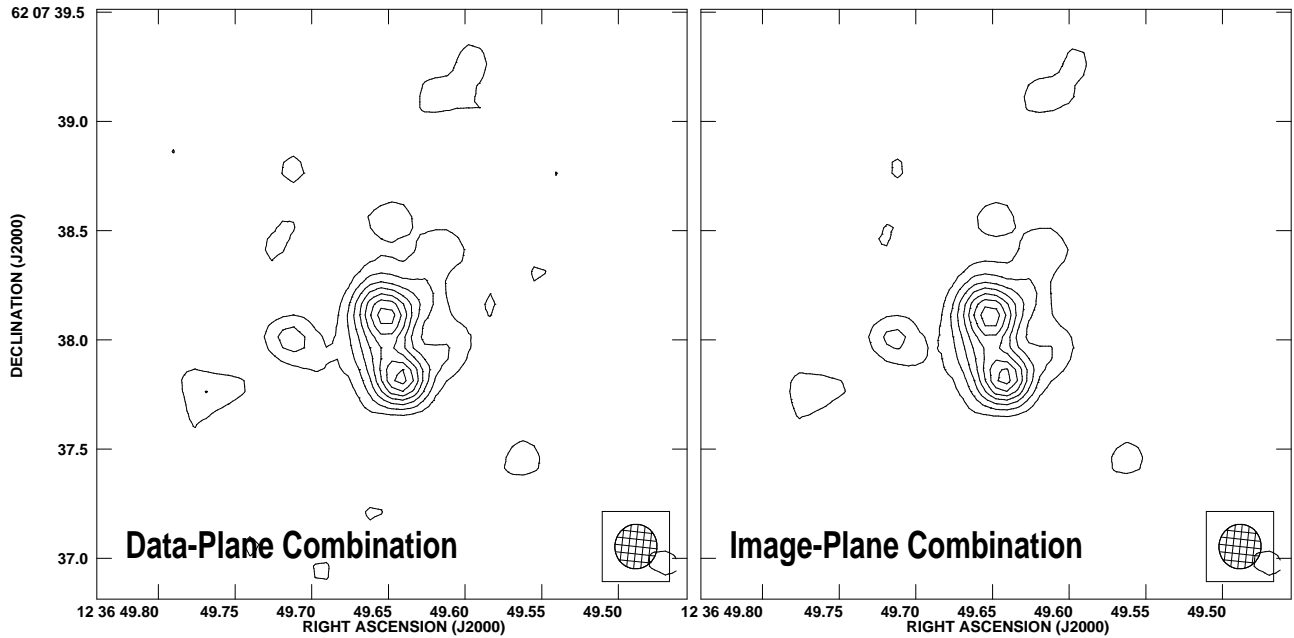}}}
\centering
\caption{A comparison of {\sc clean}ed images using MERLIN+VLA combination  in
the visibility (left) and image (right) planes. The contour interval
is linear with a lowest contour value of 10 $\mu$Jy beam$^{-1}$. }
\label{fig1} 
\end{figure*}

To demonstrate the equivalence of sky plane and visibility plane
combination, we have performed a test with one 307-$\mu$Jy source,
J123649+620738, which lies 320~arcsec almost due south from the field
centre, just outside the 10-arcmin field. The MERLIN and VLA datasets
were separately phase rotated to the position of this test source and
then averaged in time and to a single frequency channel. Averaging
restricts the field of view to a small region around the test source,
but simplifies and compresses the datasets by a considerable amount
allowing data combination prior to Fourier transformation. From the
averaged datasets, a `dirty map' and `dirty beam' were then made
and the map cleaned using the sky-plane combination method described
above. Separately, the averaged data were then combined in the
visibility plane and a second `dirty map' and `dirty beam' were
then made and {\sc clean}ed. The two {\sc clean}ed images are shown in
Fig.~\ref{fig1}; two compact components are embedded within a region
of emission extending over $\sim$1.5~arcsec.  The two images are
essentially identical; the difference between them is noise-like with
a peak difference of under 4~$\mu$Jy beam$^{-1}$ in the contoured area.

\begin{figure}
\vspace*{2cm}
\hspace{0.1cm}
\resizebox{8cm}{!}{
\epsfbox{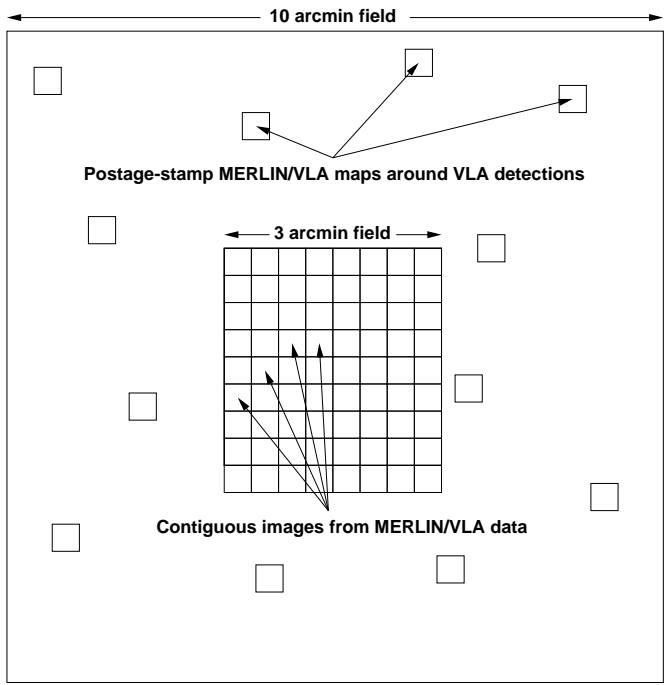}}
\centering
\caption{Schematic of the image coverage in the 3- and 10-arcmin fields. Only a few of the 92 `postage-stamp' images are indicated for simplicity.}
\label{fig2} 
\end{figure}

We also constructed a mini-mosaic of the inner 3 arcmin $\times$ 3
arcmin field (see Section~\ref{sect-3arcmin}) using the sky-plane
combination technique in order to search for additional sources below
the VLA-alone 40-$\mu$Jy detection threshold. In this case, however,
72 contiguous images were made in an $8\times9$ grid centred on the
HDF, using a pixel size of 0.0625 arcsec.  We spaced the grid so that
the central 410 -- 412 pixels (within a {\sc clean}ed area of 512 $\times$
512 pixels) of the images abutted one against another so as to cover
the field completely allowing for sky curvature. We stress that these
images are not centred on existing VLA-only detections as is the case
with the images derived for the 10-arcmin field. The image coverage of
the 3- and 10-arcmin fields is illustrated in Fig.~\ref{fig2}.

Primary beam corrections for combination images within the the
10-arcminute field are small since they are dominated by the 25-m
antennas of the VLA and MERLIN which have a half-power beam width
(HPBW) of $\sim35$ arcmin at 1.4 GHz.  The Lovell telescope has an
HPBW of 12 arcmin but on a baseline to a 25-m antenna the combined
HPBW is $\sim20$ arcmin (see Strom (2004) for the derivation).  With
respect to a source at the field centre, the flux of a source 5 arcmin
distant is only depressed by 1 per cent and 12 per cent for 25-m--25-m
and 25-m--Lovell baselines respectively.  We compared the images
from VLA-only and MERLIN$+$VLA combinations and estimated that, in the
combination images, the measured fluxes 5 arcmin from the field centre
are depressed by only $\sim6$ per cent and for the majority of sources
in the 10-arcmin field the effect is significantly less.  The position
uncertainties given in Table~\ref{10arcmin.tab} take into account the
local increase in noise due to the various aberrations and the quoted
flux densities listed in Table~\ref{10arcminchars} have been corrected
for primary beam effects. No correction has been made to the contoured
combination images shown in Appendix~\ref{sect-details}.

\subsection{Positional comparisons with previously published 1.4 GHz images}
\label{sect-pos_comp}
 Within each of the `postage-stamp' images, the sky is treated as
 flat. However, the image offset from the interferometer pointing
 centre is corrected for sky curvature so as to avoid image distortion
 at large offsets. Richards (2000) imaged the entire 40-arcmin VLA
 primary beam using 16 facets of size 14 arcmin $\times$ 14 arcmin. There is
 measurable sky distortion near the edges of these facets which
 introduce positional offsets of up to 0.2 arcsec for individual
 sources. However, after correction for these known effects, we find
 that the MERLIN and VLA positions agree to within 15 mas. The
 MERLIN+VLA combination images are of significantly greater angular
 resolution than the VLA-only images of Richards (2000). Furthermore,
 at MERLIN resolutions the vast majority of these weak radio sources
 are resolved. The positions quoted in Table~\ref{10arcmin.tab} are
 for the brightest features in the combination images, and for the
 reasons just given, differ from those quoted in Richards (2000). A
 full analysis of the astrometric alignment of this multiple
 small-image technique is given in Section~\ref{sub_roalign} where we
 find that the quoted positions for the MERLIN+VLA combination images
 in Table~\ref{10arcmin.tab} lie on the International Celestial
 Reference Frame (ICRF) to within 15 mas.

\section{THE OPTICAL IDENTIFICATIONS OF THE RADIO SOURCES}
\label{sect-align}
Williams et al.\/ (1996) describe the WFPC2 observations and data
reduction procedure and present a complete catalogue of objects detected
within the HDF. In the 4.7 arcmin$^2$ region of the HDF, objects as
faint as {\em U}~=~27.6, {\em B}~=~28.1, {\em V}~=~28.7, and {\em I} = 28.0 (AB magnitudes)
are detected at the 10 $\sigma$ level.  In addition to the HDF, eight
{\em HST} exposures of one orbit were taken in {\em I}-band (F814) in flanking
fields (HFF) immediately adjacent to the HDF (Williams et al.\/
1996).  The point source sensitivity for each of these frames is about
{\em R}~=~25 mag.

In the process of aligning the optical and radio fields, we also used
a deep CFHT {\em I}-band frame taken by Barger et al.\/ (1999). This
field is 9 arcmin on a side and encloses both the HDF and HFF
frames. The CFHT field also enabled additional optical identifications
outside the HDF and HFF to be made. This field has a resolution of
about 1.5~arcsec  and its limiting magnitude is {\em R} $\sim$26.

\subsection{Radio/optical astrometric alignment}
\label{sub_roalign}
In order to make reliable optical identifications of the radio
sources, it is important to align the {\em HST} images with the radio images which
are closely tied to the ICRF (see below). Although the fine guidance
system of the {\em HST} is accurate to a few mas, intrinsic
uncertainties in the {\em HST} Guide Star Catalog positions on the order of
1--2~arcsec are the limiting source of error in tying the {\em HST}
astrometric grid to the ICRF.

\subsubsection{Radio astrometry}
\label{sect-astrometry}
The position grid of the VLA HDF radio images is tied to the radio
source J121711+583526 with an assumed position of R.A.\/
12\h~17\m~11\fs0202 and Dec.\/ +58\degr~35\arcmin~26\farcs228 (J2000)
(Patnaik et al.\/ 1992).  This source has quoted positional errors of
13 mas with respect to the ICRF. Subsequently, the position of the
compact core of this source has been established at higher radio
frequencies to an accuracy of order 1 mas with respect to the ICRF
(Beasley et al.\/ 2001). However, the object shows significant jet
emission, especially at 1.4 GHz, and positional blending between core
and jet components result in an overall radio positional uncertainty
of $\sim$10 mas. The accuracy of transferring the position of
J121711+58585 to the position of the VLA HDF images is much better than
50 mas, derived from an estimate of typical long-term systematic phase
errors and the angular separation in the sky between J121711+58585 and
the Hubble fields.

The position of the compact MERLIN phase calibration source, which
lies less than 40 arcmin from the HDF field centre
(J124129+622041, R.\/~A.\/  12\h~41\m~29\fs589115 
Dec.\/ +60\degr~20\arcmin~41\farcs32402 (J2000), has been
established to lie on the ICRF to better than 1 mas (Beasley et
al\/ 2001). The error in transferring the position of
J124129+622041 to the position of the MERLIN images is less than 5
mas, again derived from typical systematic phase performance and the
calibration source to HDF separation.

As an internal consistency test, we compared the independently derived
MERLIN and VLA radio positions for the compact flat-spectrum AGN
source J123714+620823 (which lies some 5 arcmin from the HDF field
centre towards the edge of the 10-arcmin field); they agree to
better than 12 and 10 mas in Right Ascension and Declination
respectively. This is consistent with the quoted phase calibrator
position errors in Patnaik et al.\/ (1992). We are therefore confident
that the radio positions for the MERLIN+VLA combination images
presented in this paper lie on the ICRF to within 15 mas.

\subsubsection{Optical -- radio alignment of the CFHT field}
\label{sect-cfht_radio}
 Each {\em HST} WFPC2 frame typically contains too few radio detections to
 align the radio and optical images directly. An intermediate stage is
 thus required. We have therefore aligned the ground-based CFHT deep
 {\em I}-band image covering a 9-arcmin field centred on the HDF (Barger
 et al.\/ 1999) under the radio detections. The CFHT image has, to
 first order, already been corrected for geometrical
 distortion. However, a further correction was made in order to
 optimise the alignment of this frame with respect to the radio source
 positions and hence the ICRF. This was achieved by fitting the
 positions of 36 galaxies identified with brighter radio sources in
 the 10-arcmin radio field. Both 4- and 6-parameter fits were
 performed. The 4-parameter solution fits for an $x$ and $y$ shift, a
 rotation, and a stretch term. The 6-parameter fit allows for
 non-perpendicular $x$--$y$ axes with a further stretch term. The
 6-parameter solution was found to be no better than the 4 parameter
 solution since the two axes were found to be perpendicular to better
 than 0.01 degrees. We adopted the 4-parameter solution and applied these to
 positions derived from the CFHT frame.

\begin{figure}
\hspace{0.1cm}
\rotatebox{-90}{
\resizebox{9.5cm}{!}{
\epsfbox{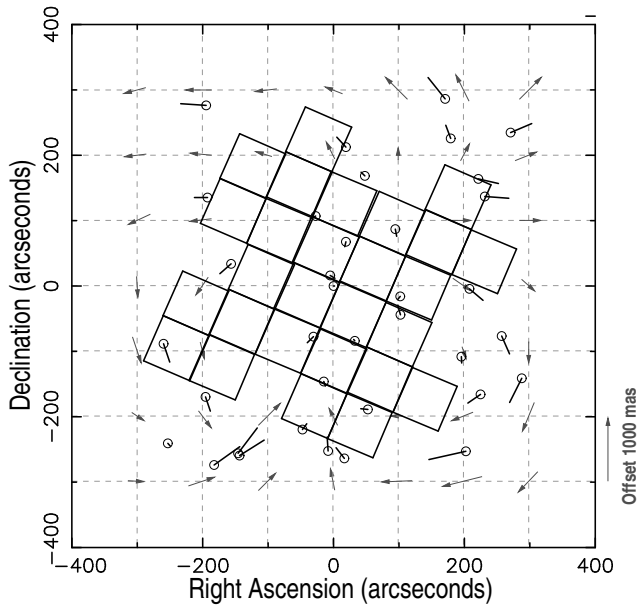}}}
\vspace*{-1cm}
\caption{Fitting residuals for the CFHT frame (Barger et al.\/
1999); the HDF and HFF WFPC2 frames are shown as solid lines. Circles
and bars mark the residual offsets from the 4-parameter fitted
solution for a number of radio sources with compact nuclear components
associated with compact optical galaxies. Arrows at grid intersections
mark the derived distortion terms used to correct galaxy positions in
the outer part of the CFHT frame after applying the 4-parameter
solutions. A 1000 mas scale for the bars and arrows is shown at bottom
right.}
\label{fig3} 
\end{figure}

The fitting residuals in the outer parts of the CFHT frame (typically
outside the area of both the HDF and HFF) show that there are small
but significant departures from the model. The fitting residuals are
shown in Fig.~\ref{fig3}. Where it was felt that the residuals showed a
consistent offset over a region in the outer parts of the CFHT frame,
a further empirical positional shift correction was made. These final
corrections are also marked in Fig.~\ref{fig3}.


\subsubsection{Alignment of CFHT and HDF/HFF WFPC2 frames}
\label{sect-cfht_hst}
After applying positional shifts from the 4 parameter solution
(derived from the whole of the CFHT frame), the rms of the fitted
residuals in the central part of the image, which overlies the 3 HDF
frames, was found to be 66 mas. For this sub-region only, a small
modification was made to the CFHT shift parameters in order to
optimise the optical to radio alignment within the HDF itself
resulting in a reduction of the rms fitted residuals to 41 mas. The
rms of the fitted residuals for the inner flanking fields (within 2.5 arcmin
 of the HDF pointing centre) was found to be 118 mas. For
the outer flanking fields the rms of the residuals was found to be 143
mas.

Optically compact galaxies in each WFPC2 frame were then registered
over the same galaxy in the positionally corrected CFHT image. By this
means, the WFPC2 frames were aligned with the radio frame and
therefore the ICRF. The optical to radio registration errors are still
the dominant source of errors in this procedure. Within the HDF
itself, the residual optical registration error is less than 50 mas,
rising to approximately 120 mas in the inner parts of the HFF, and to
approximately 150 mas at the edge of the HFF. At the edge of the CFHT
frame the registration error is $\sim$250 mas. These astrometric
errors are summarised in Table~\ref{10arcminerrs}.
\begin{table*}
\caption{Astrometric alignment errors in the 10-arcmin field.} 
\label{10arcminerrs}
\begin{tabular}{ccccc}


&&&&\\

{\bf Registration Error}  & \multicolumn{4}{c}{\bf Sub-Region of the 10-Arcmin Field} \\
 & HDF &Inner HFF & Outer HFF & Beyond HFF \\
\hline
&&&&\\
Radio $\Leftrightarrow$ ICRF \\ (MERLIN $\Leftrightarrow$ VLA $\Leftrightarrow$ ICRF) & $<$15mas & $<$15mas & $<$15mas & $<$15mas \\
&&&&\\
\hline
&&&&\\
Optical $\Leftrightarrow$ ICRF \\ (MERLIN+VLA $\Leftrightarrow$ CFHT $\Leftrightarrow$ {\em HST}) & $<$50mas & 50-100mas & 100-150mas & 150-250mas \\
&&&&\\
\hline

\end{tabular}

\end{table*}

\section{THE 10-ARCMINUTE FIELD}
\label{sect-10arcmin}
In this section we present and discuss the detailed radio structures
found for the complete sample of 92 weak radio sources stronger than
40 $\mu$Jy (5.3$\sigma$) detected in the VLA-only dataset and lying
within the 10-arcmin field centred on the HDF. To repeat, for each
source combination MERLIN+VLA `postage-stamp' images were made with
a deconvolved area 25.6$\times$25.6~arcsec$^2$ around the VLA
position. Three sizes of restoring beam were employed: the formal
fitted beam of 0.204$\times$0.193~arcsec$^2$ with major axis position
angle --6\degr, and larger circular beams of 0.3~arcsec, and
0.5~arcsec.  The complete list of source details is given in
Tables~\ref{10arcmin.tab} and~\ref{10arcminchars}, and the radio
structures at the angular resolution shown in column 5 of
Table~\ref{10arcmin.tab} are displayed in Fig.~\ref{colourpix},
overlaid on the astrometrically-aligned optical images. The optical
images are either WFPC2 HDF/HFF frames or, for those regions outside
the HFF, the CFHT frame. The additional source information found in
the central 3-arcmin field is presented in Table~\ref{3arcmin.tab} and
discussed in Section~\ref{sect-3arcmin}.  The high resolution of the
HDF/HFF observations allows us to concentrate on characterising the
phenomena responsible for radio emission; these may coexist with other
processes in the same host galaxy.

\begin{table*}
\caption{1.4-GHz radio sources detected at $\ge7\sigma$ with
0.5-arcsec resolution in the 3-arcmin field. All sources were also
detected by the VLA alone at 1.4 GHz (see Tables~\ref{10arcmin.tab}
and~\ref{10arcminchars}).}
\label{3arcmin.tab}
\begin{tabular}{lrrcccl}


&&&&&&\\

{\bf Name}  & {\bf P$_{{\bf 1.4}}$}  & {\bf $S_{{\bf 1.4}}$}&Size &P.A.&Position ({\bf P})& Other \\
&($\mu$Jy beam$^{-1}$) &($\mu$Jy)&$''\times''$&$^{\circ}$&
(J2000)& Information\\
\hline
&&&&&&\\
 J123636+621320&   29 &  50  	& (0.66$\times$0.46)$\pm$0.13 &    75$\pm$30 &12 36 36.8982  +62 13 20.320&     \\
 J123642+621331& 	348&472 & (0.339$\times$0.266)$\pm$0.001 & 103$\pm$1 &12 36 42.0959  +62 13 31.410&V$_{8.4}$I  \\
 J123644+621133&	621&791 	& (0.317$\times$0.166)$\pm$0.001 &   6$\pm$1 &12 36 44.3894  +62 11 33.110&V$_{8.4}$H      \\
 J123646+621448&   79 & 101  	& (0.425$\times$0.255)$\pm$0.001  &159$\pm$1 &12 36 46.0607  +62 14 48.728&V$_{8.4}$     \\
 J123646+621404& 	142&199 & $<0.42$$\times$$<0.42$ --- &   ---   ---       &12 36 46.3352  +62 14 04.694&V$_{8.4}$HI   \\
&&&&&&\\	        			                                             
 J123649+621313&  31 & 134  	& (1.46$\times$0.65)$\pm$0.16 &    80$\pm$6  &12 36 49.6830  +62 13 12.885&V$_{8.4}$HI   \\
 J123651+621221&  56 & 130   	& (0.84$\times$0.40)$\pm$0.07 &    88$\pm$5  &12 36 51.7238  +62 12 21.408&V$_{8.4}$HI   \\
 J123652+621444& 106  &122    	& (0.29$\times$0.19)$\pm$0.06 &   110$\pm$20 &12 36 52.8865  +62 14 44.067&V$_{8.4}$H     \\
 J123653+621139&  41  & 53    	& (0.43$\times$0.23)$\pm$0.16 &   112$\pm$19 &12 36 53.3754  +62 11 39.614&V$_{8.4}$I  \\
 J123656+621207&  27   & 30   	&$<0.64$$\times$$<0.64$  ---       &---     ---  &12 36 56.5570  +62 12 07.446&H  \\
&&&&&&\\	        			                                             
 J123701+621146&      29&130 	& (1.36$\times$0.60)$\pm$0.18 &    6$\pm$7   &12 37 01.5745  +62 11 46.738&V$_{8.4}$I   \\
\hline
\end{tabular}

\noindent
\begin{flushleft}
V$_{8.4}$ -- also detected by the VLA alone at 8.4 GHz\\
H -- radio source located in the Hubble Deep Field\\
I -- radio contours overlap the 3$\sigma$ position box of an {\em ISO}
detection\\
\end{flushleft}
\end{table*}

\subsection{Classifying the radio structures}
\label{sect-classify}
\begin{figure}
\rotatebox{-90}{
\resizebox{7.5cm}{!}{
\epsfbox{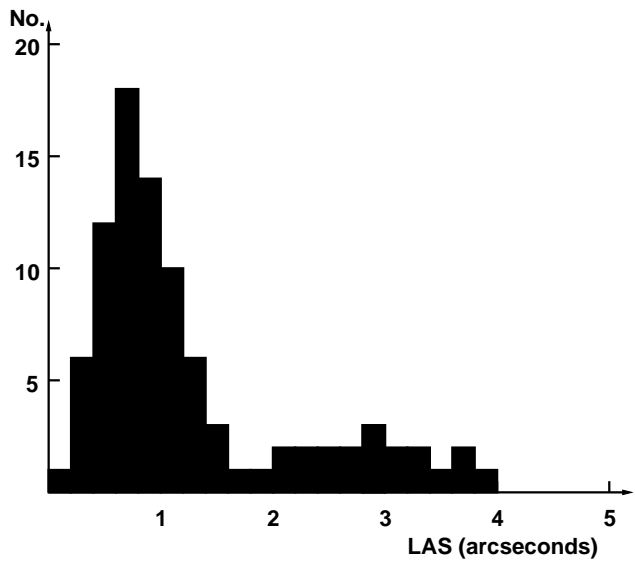}}}
\centering
\caption{Distribution of largest angular sizes for 91 of the 92 radio sources studied in the 10-arcmin field and listed in Table~\ref{10arcminchars}. The remaining source, J123644+621133, has a largest angular size of 12~arcsec.}
\label{fig4} 
\end{figure}

Although few of these $\mu$Jy radio sources are appreciably resolved
by the 2-arcsec beam of the VLA A-array image, virtually all are
resolved by the MERLIN+VLA combination showing that they have angular
sizes in the range 0.2 to 3~arcsec, typically smaller than the
sizes of the optical galaxy images. Fig.~\ref{fig4} shows the observed
distribution of angular sizes. The structural description scheme
adopted in Table~\ref{10arcminchars} is as follows:


\begin{itemize} 
\item [\bf FRI]: Fanaroff \& Riley (1974) Type I `classical' double
structure. \\

\item [\bf WAT]: Wide-Angled-Tail `classical' double
structure. \\

\item [\bf C]: Compact component at 0.2-arcsec
resolution.\\

\item [\bf C1E]: Compact component + one-sided extended
emission. \\

\item [\bf CE]: Compact component + two-sided extended
emission. \\

\item [\bf E]: Extended emission with no compact component. Note
that most of the E sources have sub-galactic dimensions. \\

\item [\bf +]: Low surface-brightness emission lies beyond that
region shown in the MERLIN/VLA image (detected by the VLA
alone). 
\end{itemize}

 There are very few radio structures typical of high luminosity
 sources (i.e. twin lobes on either side of the parent galaxy) and
 those that are found are associated with the relatively stronger,
 mJy, sources. The vast majority of $\mu$Jy sources in the 10-arcmin
 field have radio structures with sub-galactic sizes.

We have compared our 1.4-GHz flux densities with those measured by the
VLA at 8.4 GHz (Richards et al.\/ 1998) to derive radio spectral
indices for our sample (see column 5 in Table~\ref{10arcminchars} and
table 5 in Richards 2000).

The classification scheme  adopted for the radio sources is as follows:


%

\noindent {\bf AGN/AGN Candidate}: We classify a source as an `AGN'
if it has a a flat or inverted radio spectrum accompanied by a compact
core and one or two-sided extended structure\footnote{J123725+621128
has a steep spectrum at VLA resolution but the MERLIN+VLA images show
that it is unmistakably an AGN with large radio lobes}. Sources with
only some of these characteristics are classified as `AGN
candidates'.
 \vspace*{2mm}

\noindent {\bf Starburst/Starburst Candidate}: We classify a source as
a `Starburst' if it has a steep radio spectrum, is extended on
sub-galactic scales (often aligned with the galaxy major axis) and is
also detected by the Infrared Space Observatory ({\em ISO})
(Goldschmidt et al.\/ 1997; Aussel et al.\/ 1999). The correlation
between radio and IR luminosity is so close (Carilli \& Yun 1999;
Garrett 2002) that an ISO detection, especially at 15 $\mu$m, is a
strongly statistically significant indicator that at least the
proportional fraction of the radio emission is also of starburst
origin.  Sources with only some of these characteristics are
classified as `Starburst Candidates'.  Three high-redshift ($z>2$) starburst
systems which show evidence for an additional embedded AGN component,
such as a compact core or hard X-ray emission  (see
Table~\ref{10arcminchars}), are designated as S*.  Alexander et al.\/
(2002) discuss the likelihood of 15-$\mu$m sources with hard X-ray
emission harbouring AGN.
\vspace*{2mm}

\noindent {\bf Unclassified}: Sources with complex radio structures
which could be associated with either AGN or starburst activity are
listed as `unclassified', for example if the radio emission is more
extended than optical emission and there is a dearth of other
evidence.

Appendix~\ref{sect-details} gives a source-by-source summary of
detections in the 10-arcmin field.

\begin{figure*}
\resizebox{15cm}{!}{
\rotatebox{-90}{
\epsfbox{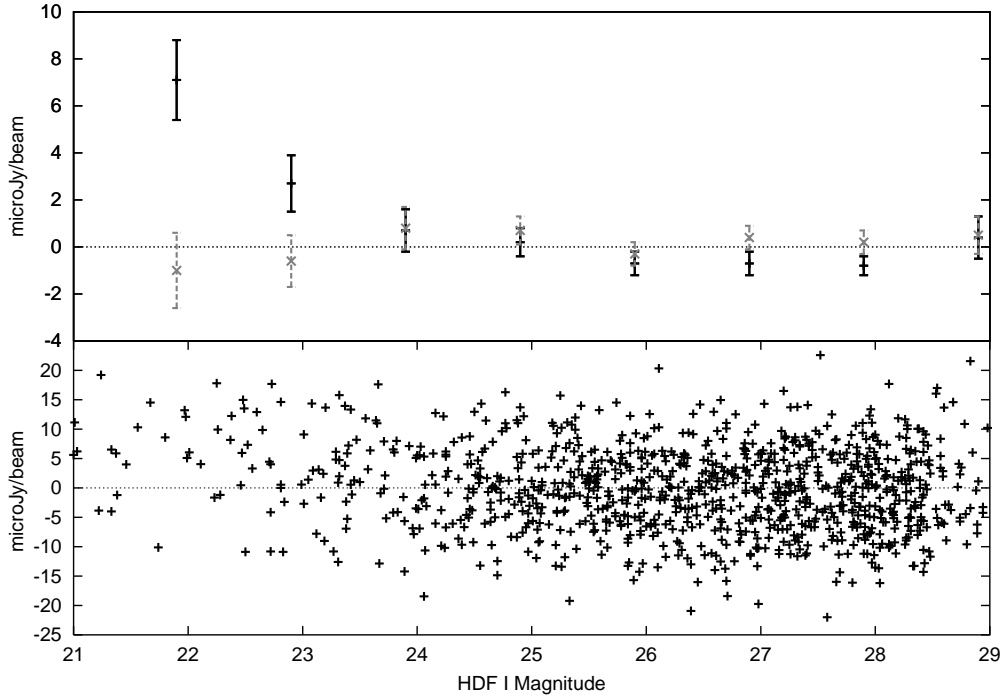}}}
\caption{(Upper) radio brightness at 1.4 GHz smoothed to 1-arcsec
resolution at the position of known galaxies, plotted as black
${\mathbf +}$ signs against the WFPC2 galaxy {\em I} magnitude, in
1-magnitude bins. The control data are displayed as grey ${\mathbf
\times}$ signs and incorporate a random 7-arcsec shift applied to the
radio positions in both R.A. and Dec.. (Lower) data for the known
galaxies shown unbinned.}
\label{fig5} 
\end{figure*}

\section{THE THREE-ARCMINUTE FIELD}
\label{sect-3arcmin}

\subsection{Mapping and analysis}

The innermost 3.41$\times$3.84 arcmin$^2$ of the radio field was
examined in more detail, and is referred to as the `3-arcmin
field'.  A grid of $8\times9$ maps was made from the calibrated
MERLIN and VLA data sets as described in Section~\ref{sect-comb}. We also restored
the {\sc clean} components for the four bright sources in this region which
had already been subtracted as described in Section~\ref{sect-merlin}.

The noise fluctuations at low brightness levels are higher than
expected for a pure Gaussian distribution. A number of factors are
likely to contribute to this excess, as discussed in Section~\ref{sect-vla}.
These include pointing fluctuations, residual ripples due to sources
in the outer parts of the primary beam and residual side-lobes of the
brighter sources in the field. Since such effects are likely to
produce as many negative deviations as positive, we have adopted a
pragmatic approach to establishing real source thresholds.

Initially each of the 72 maps was examined for regions with brightness
above 25 $\mu$Jy beam$^{-1}$; 42 of them had no such peaks.  The
brightness distribution in these `source-free' maps was
statistically analysed. The rms noise level with a 0.5-arcsec
restoring beam is 3.9 $\mu$Jy beam$^{-1}$ with no positive or negative
peaks above $\pm 7\sigma_{\rm rms}$.  The noise level with the
smallest 0.2-arcsec restoring beam is 3.3 $\mu$Jy beam$^{-1}$ with no
peaks above $\pm 8\sigma_{\rm rms}$. In the other 30 images, positive
regions above these levels (27 and 25 $\mu$Jy beam$^{-1}$ at 0.5- and
0.2-arcsec resolutions respectively) are therefore considered reliable
detections.


Eleven positions have peaks above 27 $\mu$Jy beam$^{-1}$ (see
Table~\ref{3arcmin.tab}). All these sources coincide with VLA 1.4-GHz
detections with flux densities $>40$ $\mu$Jy (total); they can all be
regarded as secure and thus the sample is complete to this limit. The
positions and sizes of these eleven sources were found by fitting
Gaussian components in the image plane.  Nine sources are
significantly resolved, and one (J123656+621207) is too weak for its
size to be measured reliably. One object in Table~\ref{10arcmin.tab}
(J123646+621445), which also lies in the 3-arcmin field, is so
spatially resolved as to fall below the 27 $\mu$Jy beam$^{-1}$
detection threshold and thus does not appear in
Table~\ref{3arcmin.tab}. This imaging exercise therefore revealed no
new sources in the 3-arcmin field in the brightness range 27--40
$\mu$Jy beam$^{-1}$. Note however that at the same time as reducing
the detection threshold from 40 (VLA-only) to 27 $\mu$Jy beam$^{-1}$,
we also increased the angular resolution from 2~arcsec (VLA-only) to
0.5~arcsec. The slope of the integral source counts around 40 $\mu$Jy
is --1.4 (Richards 2000) and one would therefore expect to find
$\sim$8 additional sources within the 3-arcmin field in the flux
density range 27--40 $\mu$Jy, including one compact AGN. Since we do
not detect any new sources, we infer that most objects weaker than
40~$\mu$Jy are heavily resolved with a 0.5-arcsec beam, do not contain
compact radio components $>27$~$\mu$Jy and must have angular sizes
$>1$ arcsec. We ascribe the failure to detect an additional compact
AGN source to chance. The planned extension of the investigation to
the complete mapping of the 10-arcmin field (see
Section~\ref{sect-future-radio-obs}) will provide additional
constraints on this interpretation.

\subsection{Statistical comparison with HDF galaxies}
\label{sect-stat}
For the part of the 3-arcmin field which overlies the central HDF area
itself, the MERLIN+VLA images were smoothed to 1-arcsec resolution,
after excluding all 11 significant individual source detections
(points above +27 $\mu$Jy beam$^{-1}$). The radio brightness was then
measured at the positions of all catalogued HDF galaxies (Williams et
al.\/ 1996). The results are shown in Fig.~\ref{fig5}, binned by {\em
I}-band magnitude. The figure shows that sources over an order of
magnitude weaker than our detection threshold (40 $\mu$Jy), are
statistically associated with galaxies brighter than {\em I} $\sim23$
mag. A control sample which incorporates a random 7-arcsec shift
applied to the radio measurement positions, in both R.\/A. and Dec.\/ with respect to the optical
galaxies shows no excess of detected brightness. Thus the source
population down to a few $\mu$Jy shares similar properties with the
10-arcmin field sample, being identified statistically with the
brighter HDF galaxies.



\begin{figure*}
\vspace*{0cm}
\hspace*{-1cm}
\rotatebox{-90}{
\resizebox{10cm}{!}{
\epsfbox{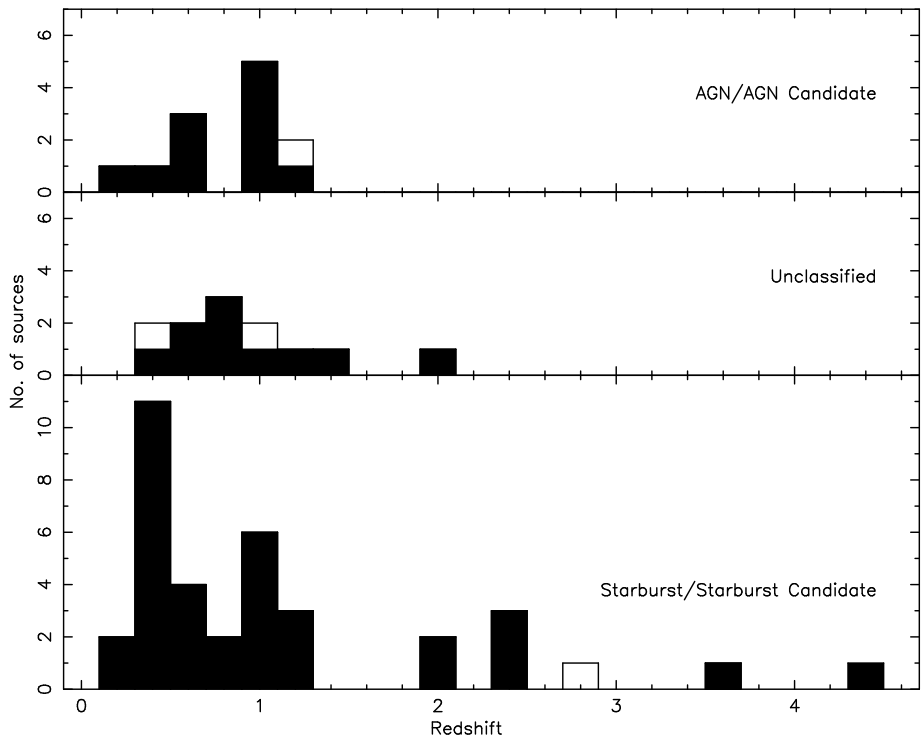}}}
\vspace*{1cm}
\caption{Redshift distribution for those sources in the 10-arcmin field with measured spectroscopic redshifts (black) and estimated photometric redshifts (white).}
\label{red} 
\end{figure*}  

\begin{figure*}
\vspace*{0cm}
\hspace*{-1cm}
\rotatebox{-90}{
\resizebox{6cm}{!}{
\epsfbox{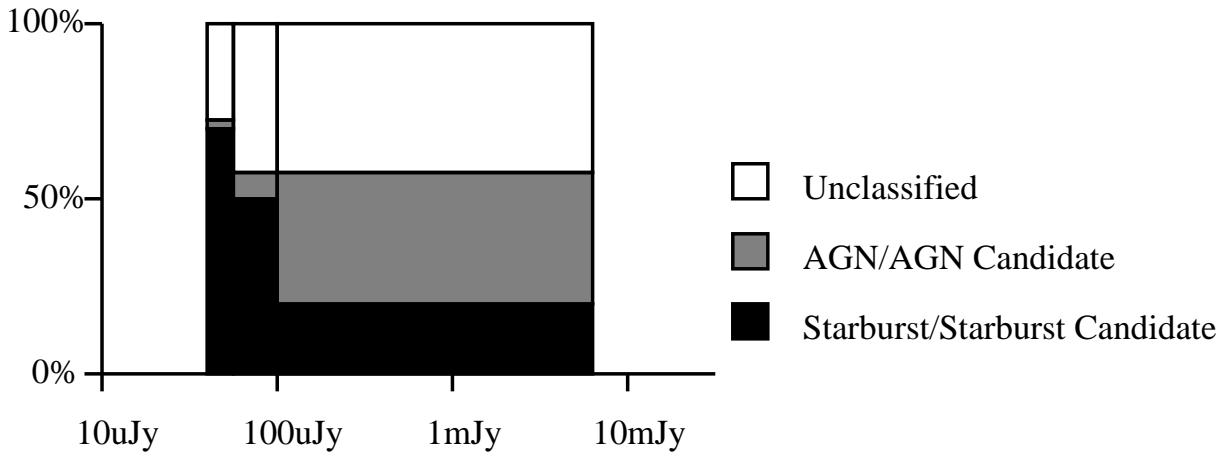}}}
\vspace*{1cm}
\caption{Distribution of source classifications for the objects in the
10-arcmin field in each of three flux ranges (40--55 $\mu$Jy, 56--105 $\mu$Jy, \& 106--6000 $\mu$Jy) containing 30, 30, \& 32 sources
respectively.}
\label{sourceclass} 
\end{figure*}

\section{PROPERTIES OF THE RADIO SOURCES}
\label{sect-properties}

\subsection{Source sizes}
\label{sect-sizes}
As summarised in Fig.~\ref{fig4}, the great majority of the $\mu$Jy
source population have radio angular sizes in the range 0.2~arcsec to
3~arcsec. Only one source in the sample of 92 objects is unresolved
(J123714+620823). Only two sources (J123644+621133 and J123725+621128)
show classical radio structures usually associated with high
luminosity AGN, and both of these are representative of the tail of
the mJy source population rather than that of the $\mu$Jy sources. One
other AGN system, (J123652+621444), shows evidence from WSRT
observations of a possible one-sided low surface-brightness jet
structure extending for some 30~arcsec with an implied size of around
100 kpc.

Notwithstanding that perhaps a total of 10~per cent of low-surface
brightness extended sources have been missed by the original VLA-only
survey (see Section~\ref{sect-wsrt}), the great majority of objects in the
10-arcmin field have radio structures extended predominantly on
galactic and sub-galactic scales. Their radio structures have been
defined and discussed in Section~\ref{sect-classify}
and Appendix~\ref{sect-details} respectively.  The typical source angular scale
appears similar down to 27 $\mu$Jy and even beyond (see
Section~\ref{sect-3arcmin}).

The fact that the typical angular size of $\mu$Jy sources is
$\sim$1~arcsec may have implications for future deep surveys with the
Square Kilometre Array (SKA) which could encounter a natural confusion
limit. If the slope of the source counts at 1.4 GHz remains as steep
as $-1.4$ then surveys, with limiting flux densities 100 times fainter
than the present one, will reach surface densities $>$500 sources
arcmin$^{-2}$ (similar to the surface density of optical galaxies in
the HDF). If the typical source size remains $\sim $1~arcsec then the
proportion of sky covered with radio sources, $C$, will be $>$0.14. If
$C > 0.05$ significant blending of sources begins to occur (Fomalont
et al.\/ 2002) and obviously if $C > 0.14$ a large amount of source
blending would result, regardless of the size of the synthesized
observing beam.  This simple inference depends, however, on two large
extrapolations (source counts and sizes). The optical HDF is not
significantly confused and it could be, for example, that the fainter
radio sources are smaller, being associated with the fainter, smaller,
irregular galaxies; this could reduce $C$ to below the natural
confusion limit.

\subsection{Optical identifications and redshifts}
\label{sect-z}
Sixty-one of the 92 sources (65~per cent) in the 10-arcmin field have
measured spectroscopic or estimated photometric redshifts (at the time
of performing this analysis). Fig.~\ref{red} shows the redshift
distribution for each of the three classifications. Note that the
highest redshift source, J123642+621331, appears to contain a compact
AGN. It is clear that it is only the starburst systems which extend to
redshifts in excess of 2. For those systems with redshifts less than
2, we have compared the redshift distributions for each
classification.

The AGN and unclassified systems have median redshifts of 0.91 and
0.85 respectively, and a Kolmogorov-Smirnov goodness of fit test shows
that there is a 79 per cent probability that they are drawn from the
same population. Conversely, the starburst systems have a
significantly lower median redshift of 0.56 and there is only a 7 per
cent probability that they are drawn from the same population as the
AGN systems. Thus it appears that majority of the starburst systems
are associated with galaxies at a lower redshift than the AGN
sources. On this basis, the majority of the unclassified sources are
most likely to be AGN dominated systems with some starbursts
included. It is worth noting, however, that nearly half the starburst
systems appear in two adjacent histogram bins covering $0.3\le z<0.7$.
The excess of sources in these bins is
correlated with the redshifts of two major clusters found by Cohen et
al.\/ (2000) in their Caltech Faint Galaxy Redshift Survey of the HDF.

The histogram of starburst galaxies and starburst candidates also
shows evidence for a tail extending to redshifts in excess of 4
(although the highest redshift candidate may also contain an
AGN). Fig.~\ref{SB} shows that many of these high redshift starburst
systems have been identified as sub-mm sources (Chapman et al.\/
2004a, Chapman  et al.\/
2004c submitted). In addition, two are optically
faint and extremely red, suggesting that they are dust-enshrouded. Of
the thirteen optically faint radio sources in the 10-arcmin field,
only two are classified as AGN or AGN candidates. Thus, although the
majority of starburst systems are associated with galaxies brighter
than {\em I}~=~25 and at redshifts less than two, it appears that we
are beginning to sample an additional population of high redshift
systems. In addition, some of these high redshift systems may be
complex, containing both dust-shrouded starbursts and embedded AGN
(see Appendix~\ref{sect-details} and Section~\ref{sect-faint}).

\subsection{Structural classification and source flux density}
\label{sect-struct}
Fig.~\ref{sourceclass} summarises the classification of the 92 sources
in the 10-arcmin field in each of three flux density ranges (40--55
$\mu$Jy, 56--105 $\mu$Jy, and 106--6000 $\mu$Jy); these contain
approximately equal numbers of sources: 30, 30, and 32
respectively. It is immediately apparent that the proportion of
starburst and starburst candidate systems rapidly increases with
decreasing flux density below 100 $\mu$Jy. In the highest flux density
range, there are only around 20~per cent starburst systems with around
40~per cent AGN and 40~per cent unclassified sources. In the lowest flux
density range, the proportion of starburst systems has risen above
70~per cent while the proportion of AGN systems has dropped to less
than 5~per cent. Fomalont et al.\/ (2002) note that in their sample of
radio sources at 8.4 GHz, the average radio spectrum for those sources
identified with optical counterparts becomes steeper for fainter
sources ($S_{8.4}<$35 $\mu$Jy). These authors also state that objects
fainter than {\em I}~=~25.5 mag predominantly have steep radio
spectra. Thus it is clear that at flux densities below around 50
$\mu$Jy at centimetric wavelengths, a population of steep radio
spectrum starburst systems begins to dominate.

There is, however, a selection effect which could reduce the numbers
of AGN identified at low flux levels. AGN systems are easily
identified if they have flat or inverted radio spectra. Radio spectral
information is derived from the present 1.4-GHz observations and those
at 8.4 GHz (Richards et al.\/ 1998). The 8.4-GHz image is extremely
sensitive (rms noise~=~2.78 $\mu$Jy beam$^{-1}$); however the small
primary beam of the VLA antennas (HPBW~=~5.2 arcmin) restricts the
region of source detection to a field of around 8 arcmin in
diameter. Weak sources close to the edge of the 10-arcmin field will
not be detected at 8.4 GHz, and this is likely to prevent the correct
identification of weak AGNs. Such systems are likely then to be
designated as `unclassified'. However, since less than 20 per cent of the
unclassified sources are found in the lowest flux range shown in
Fig.~\ref{sourceclass}, it is thought that few if any AGN have been
mis-identified at low flux levels. Moreover the {\em K}-correction
works in favour of detecting AGN in comparison with starbursts at high
$z$, as noted in Section~\ref{sect-iso}.

\begin{figure}
\vspace*{-3cm}
\hspace*{-1cm}
\rotatebox{-90}{
\resizebox{10cm}{!}{
\epsfbox{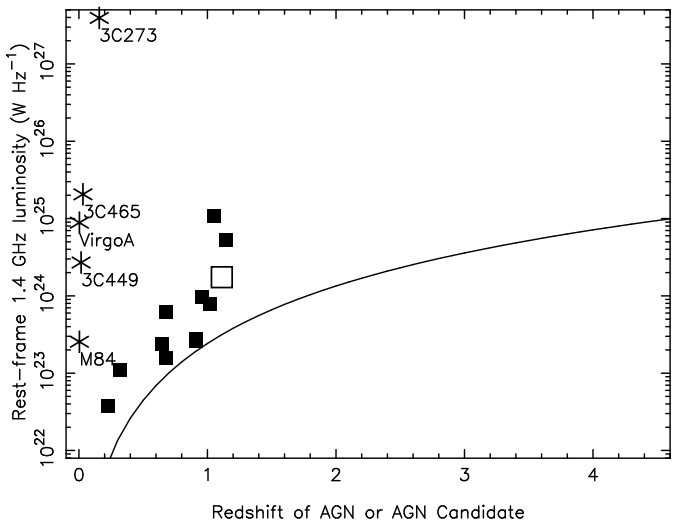}}}
\caption{1.4-GHz monochromatic luminosity for AGN and AGN candidate
sources with redshifts in the 10-arcmin field. Filled symbols are
spectroscopic redshifts, open symbols are photometric redshifts.
Asterisks show named sources. The plotted line marks the detection
threshold of 40 $\mu$Jy for sources assuming a spectral index of 0
($\Omega_{\rm V}~=~0.7$, $\Omega_{\rm M}~=~0.3$, and H$_0~=~65$
km~s$^{-1}$~Mpc$^{-1}$).}
\label{AGN} 
\end{figure}  
\begin{figure}
\vspace*{-3cm}
\hspace*{-1cm}
\rotatebox{-90}{
\resizebox{10cm}{!}{
\epsfbox{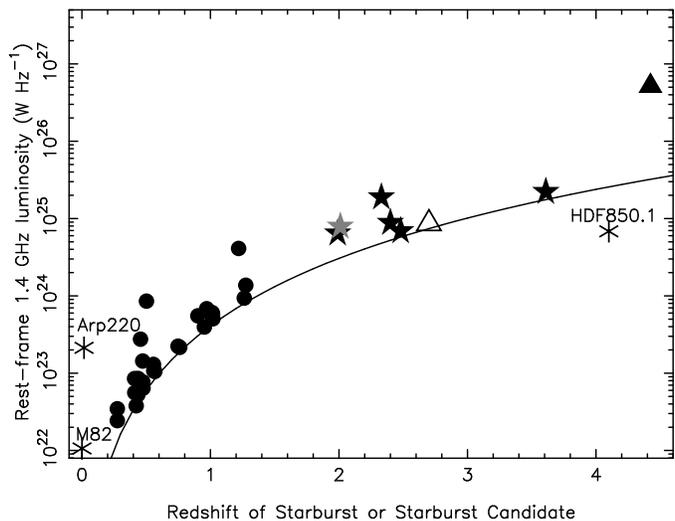}}}
\caption{1.4-GHz monochromatic luminosity for starburst and starburst
candidate sources with redshifts in the 10-arcmin field. Filled
symbols are spectroscopic redshifts, open symbols are photometric
redshifts. Five-pointed stars mark sub-mm systems. Triangles mark two
high redshift systems with possible embedded AGN, (J123642+621331, \&
J123651+621221). The sub-mm system with a possible embedded AGN
(J123635+621424) is shown in grey. Asterisks show named sources. The
plotted line marks the detection threshold of 40 $\mu$Jy for sources
assuming a spectral index of 0.7 ($\Omega_{\rm V}$~=~0.7, $\Omega_{\rm
M}$~=~0.3, and H$_0$~=~65 km~s$^{-1}$~Mpc$^{-1}$).}
\label{SB} 
\end{figure}

\subsection{Source luminosities and star-formation rates}
\label{sect-sfr}
Figs.~\ref{AGN} and~\ref{SB} show the rest-frame 1.4-GHz monochromatic
luminosities for the AGN, and starburst/starburst candidate sources in
the 10-arcmin field with measured or estimated redshifts respectively,
together with a number of well-known lower redshift radio sources of
each type. The detection threshold for 40 $\mu$Jy is plotted as a line
assuming a radio spectral index of 0 for the AGNs and 0.7 for the
starburst systems.

All the AGN candidates are well above the detection threshold which
simply reflects the fact that they have flux densities well above 40
$\mu$Jy. Their luminosities are similar to those of low luminosity,
low redshift FRI type radio galaxies like M84, Virgo A, and 3C449
(Fanaroff \& Riley, 1974); but significantly lower than those of FRII
type sources and powerful quasars like 3C273 (see Fig.~\ref{AGN} where
the source 3C465 has a rest-frame luminosity intermediate between FRI
and FRII type systems). However, apart from two sources
(J123644+621133 and J123725+621128) which show `classical' double
radio structures, all the remaining AGN systems are small core-jet
systems extended on sub-galactic scales.

The starburst candidates lie closer to the detection threshold since
they represent the majority of the weaker radio sources in the
10-arcmin field. All the systems are substantially more powerful than
M82 and about half are more luminous than Arp 220. The sources at
$z>2$ (marked as stars and triangles in Fig.~\ref{SB}) have
particularly high luminosities exceeding that of Arp 220 by an order
of magnitude.  The object at the highest redshift (J123642+621331)
which is more than 3 orders of magnitude more luminous than Arp 220,
may be a hybrid system containing both a dust-shrouded starburst and
an embedded AGN (see discussion in Appendix~\ref{sect-details}). The
other two starburst sources which appear to contain compact AGN are not
noticeably super-luminous.  Star-formation rates for starburst and
starburst candidate systems with measured redshifts are shown in
Table~\ref{sfr.tab}. These are derived from the rest-frame
monochromatic 1.4-GHz luminosity via the relationship derived by Condon
\& Yin (1990) and Condon (1992). Star-formation rates for stellar
masses $>5$ M$_{\odot}$ were calculated using the relationships given
by Cram et al.\/ (1998) and then multiplied by 5.5 to give the total
star-formation rate for stellar masses between 0.1 M$_{\odot}$ and 100
M$_{\odot}$ using a Salpeter IMF.

\begin{table*}
\caption{Rest-frame monochromatic 1.4-GHz Luminosities and star-formation rates
(assuming Salpeter IMF, see Section~\ref{sect-sfr}) for starburst systems with
measured or estimated redshifts. Star-formation rates marked with an *
may be overestimates as these galaxies are thought to contain an
embedded AGN.}
\label{sfr.tab}
\begin{tabular}{cccc}


&&&\\

{\bf Name}  & {\bf L$_{{\bf 1.4}}$}  & Redshift & Approximate Star-formation Rate\\
            &      (W Hz$^{-1}$)      &          &  (M$_{\odot}$ yr$^{-1}$)\\
\hline
&&&\\
 J123606+621021  &     $1.9\times10^{25}$    & 2.33     &     4400 \\
 J123607+621328  &     $5.2\times10^{22}$    & 0.4353   &     12 \\
 J123608+621553  &     $7.0\times10^{22}$    & 0.4593   &     16 \\
 J123612+621138  &     $2.4\times10^{22}$    & 0.275    &     6 \\
 J123612+621140  &     $3.5\times10^{22}$    & 0.275    &     8 \\
 J123615+620946  &     $9.4\times10^{23}$    & 1.263    &     220 \\
&&&\\
 J123616+621513  &     $2.2\times10^{25}$    & 3.61     &     5100 \\
 J123619+621252  &     $1.4\times10^{23}$    & 0.473    &     34 \\
 J123621+621109  &     $6.1\times10^{23}$    & 1.014    &     140 \\
 J123622+621629  &     $8.7\times10^{24}$    & 2.40     &     2000 \\    
 J123622+620945  &     $2.2\times10^{23}$    & 0.7479   &     42 \\
&&&\\
 J123630+620923  &     $4.0\times10^{23}$    & 0.953    &     93 \\
 J123632+621700  &     $8.5\times10^{22}$    & 0.437    &     20 \\
 J123633+621005  &     $5.8\times10^{23}$    & 1.016    &     135 \\
 J123634+621213  &     $2.7\times10^{23}$    & 0.456    &     64 \\
 J123634+621241  &     $4.1\times10^{24}$    & 1.219    &     960 \\
&&&\\
 J123635+621424  &     $7.8\times10^{24}$    & 2.011    &     1800 \\
 J123642+621331  &     $5.2\times10^{26}$    & 4.424    &     120000* \\
 J123646+621629  &     $8.5\times10^{23}$    & 0.502    &     200 \\
 J123646+620833  &     $6.8\times10^{23}$    & 0.9712   &     160 \\
 J123649+621313  &     $6.4\times10^{22}$    & 0.475    &     15 \\
&&&\\
 J123650+620801  &     $1.1\times10^{23}$    & 0.559    &     25 \\
 J123650+620844  &     $8.1\times10^{22}$    & 0.434    &     19 \\
 J123651+621030  &     $8.5\times10^{22}$    & 0.410    &     20 \\
 J123651+621221  &     $8.6\times10^{24}$    & 2.7      &     2000* \\
 J123653+621139  &     $1.4\times10^{24}$    & 1.275    &     320 \\
&&&\\
 J123656+621301  &     $7.7\times10^{22}$    & 0.474    &     18 \\
 J123659+621449  &     $2.2\times10^{23}$    & 0.761    &     51 \\
 J123705+621153  &     $5.5\times10^{23}$    & 0.902    &     130 \\
 J123707+621408  &     $6.8\times10^{24}$    & 2.48     &     1600 \\
 J123708+621056  &     $3.8\times10^{22}$    & 0.422    &     9 \\
&&&\\
 J123711+621325  &     $6.4\times10^{24}$    & 1.99     &     1500 \\
 J123714+621558  &     $1.1\times10^{23}$    & 0.567    &     25 \\
 J123716+621643  &     $1.3\times10^{23}$    & 0.557    &     30 \\
 J123716+621007  &     $5.6\times10^{22}$    & 0.411    &     13 \\
 J123721+621346  &     $5.0\times10^{23}$    & 1.019    &     120\\
&&&\\
\hline
\end{tabular}

\noindent
\begin{flushleft}
\end{flushleft}
\end{table*}

\section{COMPARISON WITH OTHER CATALOGUES}
\label{sect-other}

\subsection{The WSRT catalogue}
\label{sect-wsrt}
After our data had been taken and analysed, it has become clear that a
number of extended, low surface-brightness radio sources may have been
missed by the initial VLA-only finding survey. Using the WSRT
interferometer, which has much lower angular resolution (15~arcsec)
and is thus more sensitive to extended objects than the VLA A-array
(2-arcsec resolution), Garrett et al.\/ (2000) detect additional
radio sources in the region of the 10-arcmin field. Furthermore,
they identify at least two (J123720+621247 and J123636+621132) with
nearby, extended star-forming galaxies. Since these objects were not
detected by the VLA, it implies they are extended on the 5 -- 20~arcsec
angular scale and do not contain compact radio structure. However, the
crowded nature of the optical field makes identification for most of
these new detections very uncertain. In addition, Garrett et al.\/
(2000) state that further analysis of the full field is required to
distinguish how many of the WSRT-only detections are real and
associated with discrete, extended radio sources, and how many are
blends of closely-separated, faint radio sources.

The WSRT source catalogue given by Garrett et al.\/ (2000) lists 85
objects above a 5$\sigma$ limit of $\sim$50 $\mu$Jy within the
10-arcmin field. Of these 63 are also detected by the VLA hence 22 are
WSRT-only detections. The VLA-only flux densities listed in
Table~\ref{10arcminchars} include 78 sources above 50 $\mu$Jy
(6.7$\sigma$) of which 15 are VLA-only detections. One possible
explanation is that these sources have varied between the epochs of
the VLA and WSRT observations. However, many of the VLA-only
detections show extended radio structures in the MERLIN+VLA
combination images and have steep radio spectra -- indeed, in
Section~\ref{sect-properties}, we argue that the majority of weakest
radio sources in the 10-arcmin field are extended starburst
systems. These objects are very unlikely to be variable radio
sources. The implication is that the WSRT catalogue is unreliable at
this level, and that of the 22 WSRT-only detections, perhaps only
between 8 and 10 are real extended low-surface brightness radio
sources lying within the 10-arcmin field which have been missed by the
VLA. It is still likely, therefore, that the VLA-based source list in
Table~\ref{10arcminchars} is incomplete and is missing $\sim$10~per
cent of the $\mu$Jy radio source population.

\subsection{{\em ISO}}
\label{sect-iso}

Aussel et al.\/ (1999) detected 100 sources at 6.7 or 15 $\mu$m or
both within the HDF/HFF.  The {\em ISO} beams were 3 and 6 arcsec
within $\sim3.5$ and $\sim5$ arcmin fields of view at the shorter and
longer wavelengths respectively; 28 radio sources lie within the
maximum {\em ISO} field of view.  Of these, 15 have IR counterparts
within 3 arcsec.  In addition the core of the FRI J123644+621133 is
just over 3 arcsec from a 6.7-$\mu$m source, found only by the {\em
ISO} detection method described by Goldschmidt et al. (1997).  Two
more (J123646+621445 and J123656+621301) are at separations of 3--6
arcsec; in the case of J123646+621445 the IR source appears to be
associated with a separate nearby galaxy (see
Appendix~\ref{sect-details}). Note that in the column in
Table~\ref{10arcminchars} labelled {\em ISO}, `--' signifies that
the source is outside the {\em ISO} field and `No' signifies a
non-detection within the field at the local flux-density limit, see
Aussel et al.\/ (1999) for details of how this varies with position
and frequency.  Twelve of the radio sources with IR counterparts are
classified as starburst galaxies using the criteria in
Section~\ref{sect-classify} (of which 3 have embedded AGN), two are
AGN and two are unclassified. The majority of radio sources (7) within
the {\em ISO} field of view which do not have IR counterparts are also
starburst candidates, 3 are unclassified and 2 are AGN/AGN candidates.

 Starbursts typically have a steep spectrum in both the IR and the
radio whilst AGN have a flatter radio spectrum. Hence the {\em
K}-correction implies that, at the same luminosity, AGN should be
detectable at greater distances than starbursts and the proportion of
the latter may be underestimated.  Elbaz et al. (2002) showed that 80
-- 90 per cent of the extragalactic 15 $\mu$m background is of
starburst origin and 3/4 of ISOCAM galaxies are ULIRGs (Ultra-Luminous
IR Galaxies). The proportions are consistent in other deep surveys
such as of the Lockman Hole as well as in the HDF.

Garrett (2002) demonstrated that the 1.4-GHz and 15-$\mu$m rest-frame
luminosities of HDF sources detected by both WSRT and {\em ISO} are
closely correlated out to at least $z\sim1.3$, suggesting a common
emission environment. This correlation is based on using a
starburst-model spectral index in the IR and the vast majority of the
radio sources with associated IR emission are independently classified
as starburst candidates. A third of the radio sources in the sample
identified by Garrett (2002) are among the non-detections seemingly
resolved-out in the MERLIN+VLA data (Section~\ref{sect-wsrt}). One may
be a misidentification and 10 out of the 12 remaining sources have
extended emission and other starburst characteristics.

\begin{figure}
\vspace*{-5cm}
\hspace*{-1cm}
\rotatebox{-90}{
\resizebox{10cm}{!}{
\epsfbox{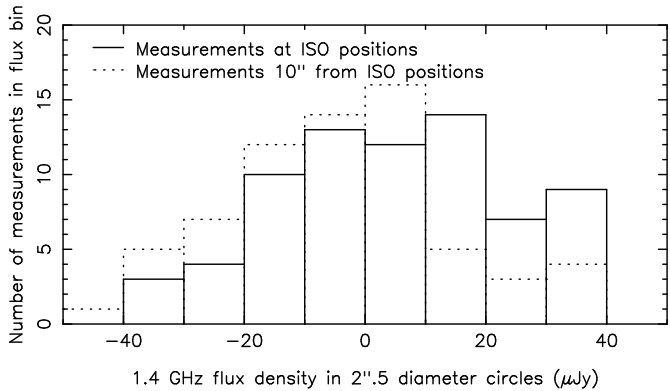}}}
\caption{Histogram showing the MERLIN+VLA 1.4-GHz brightness at
the positions of {\em ISO} detections within the $3'$ field, compared with
the off-source brightness. The `on source' histogram is shifted to
the right of the `off source' one indicating a weak statistical
association of radio emission with {\em ISO} sources.}
\label{ISOMHIST1.25.PS} 
\end{figure}

73 {\em ISO} sources lie within the 3-arcmin field.  In order to
investigate statistical associations with the {\em ISO} sources, the
mean MERLIN+VLA flux density from within concentric circles centred on
the {\em ISO} positions was measured, together with equivalent flux
densities of a control sample incorporating a 10-arcsec random
position offset from the radio position. The most significant
correlation with the radio brightness is obtained with circles of
2.5~arcsec diameter. The results are shown in
Fig.~\ref{ISOMHIST1.25.PS} and we note that radio flux densities $>10$
$\mu$Jy were found at the positions of 30 {\em ISO} detections, about
half of which are probably significant statistically.  Although this
is only a $\sim$2$\sigma$ result it indicates that deeper surveys and
more accurate IR positions, as will be obtained with the {\em Spitzer}
satellite, are likely lead to many more radio-IR associations.

\subsection{{\em Chandra}}
\label{sect-chandra}

43 of the 92 sources in the 10-arcmin field sample (47~per cent) were
detected in X-rays by {\em Chandra} using a 1 Ms exposure
(Hornschemeier et al.\/ 2001). A further 12 counterparts with
full-band counts $<65$ were found in the full 2 Ms {\em Chandra}
exposure recently catalogued by Alexander et al.\/ (2003).  These are
all listed in Table~\ref{10arcminchars}.  Note that in the column
labelled {\em Chandra}, `No' signifies a non-detection within the
field at the local flux-density limit, see Alexander et al.\/ (2003)
for details of how this varies with position and energy.  The {\em
Chandra} field of view at high sensitivity is slightly larger than the
10-arcmin radio field.  

We note that the X-ray detection rate appears
to be uncorrelated with the radio source flux density since in each of
the three flux ranges shown in Fig.~\ref{sourceclass}, 16/30, 20/30,
and 19/32 radio sources are detected by {\em Chandra} respectively. In
addition, Fomalont et al.\/ (2002) state that the seven X-ray
identifications (from the 1 Ms observation) in the HDF itself show no
trends in radio spectral index or optical counterpart and are
identified with galaxies of various spectral types. There is no
obvious correlation between X-ray and radio flux densities or
classification. Since Fig.~\ref{sourceclass} shows that source
classification is a strong function of radio flux density, this
implies that the X-ray emission from the $\mu$Jy radio source
population is not a good predictor of AGN activity in the HDF unless
this arises from a separate mechanism within the same galaxy.

A long series of papers have appeared analysing the {\em Chandra}
results using multi-wavelength comparisons.  However thus far most
considerations of radio data have used the 1 Ms X-ray catalogue and
the VLA-only radio data.  In the next paper we intend to use Virtual
Observatory tools to compare the new more sensitive samples which, for
the first time, contain enough sources in common to allow comparisons
based on selection by specific properties such as spectral index or
redshift.

\subsection{SCUBA sub-mm sources}
\label{sect-scuba}
Identifying the optical/NIR counterparts to sub-mm sources has proved
difficult due to the poor spatial resolution and astrometric
capability of current sub-mm telescopes. However, high resolution
radio images offer new possibilities for locating sub-mm sources and
for the determination of their properties. For starburst systems, it
is well established that the radio continuum flux (due to
synchrotron emission from SNR and violent turbulence associated with
intense star-formation) and the FIR/sub-mm flux (due to dust from
stellar winds) are tightly correlated as a result of their both being
linearly related to the massive star-formation rate (e.g. Helou et
al.\/ 1985, Helou \& Bicay, 1993). If the FIR-radio correlation
extends to high redshift then in principle, sensitive radio
observations can be used to pinpoint distant sub-mm sources.

From radio--far-infrared SED-based redshift arguments Smail et
al.\/  (2000) have suggested that the median redshift for the bright
SCUBA galaxy population is $\sim$3, with the substantial majority of
SCUBA sources lying at redshifts $>2$.  Dunlop et al.\/ (2004)
argue that since SCUBA sources in other fields have been identified
with relative ease compared with HDF850.1 (see below), the majority
probably lie at redshifts $<4$.  This is in agreement with Hughes 
et al.\/  (1998) who have assumed a redshift range of $2<z<4$ for
calculating the contribution made by sub-mm sources to star-formation
density. Ivison et al.\/ (2002) have investigated the SCUBA source
population in the Lockman Hole and ELAIS N2 regions with deep
multi-band imaging. With good radio positional information, they
identify radio counterparts for 18 of 30 sub-mm sources.  Of these,
they find that at least 60~per cent of the host galaxies appear to be
morphologically distorted, suggesting that the radio and sub-mm
emission arises from extended starbursts and that interactions are
common.

With all this in mind, and armed with our accurate radio-optical
astrometry, we have revisited the issue of sub-mm source
identifications in the HDF and HFF. We have used the results of
Serjeant et al.\/ (2003) who have re-measured the positions for the
eight brightest sub-mm sources in a field centred on the HDF (Hughes
et al.\/ 1998). Following discussion with S. Serjeant on the
sub-mm position errors, we have drawn error circles with a radius of
2$\sigma$ where $\sigma=\sqrt{1^{2}+{\theta_{\rm SNR}}^{2}}$
arcsec. In this quadratic combination the first term is an estimate of
the positional error due to confusion and the second is the standard
random error due to finite signal-to-noise ratio ($\theta_{\rm
SNR}=0.5 \times$ FWHM/SNR). FWHM is the size of the imaging
beam. Fig~\ref{scubapix} shows these error circles overlaid on the
astrometrically aligned optical fields. The combined MERLIN/VLA radio
images at the position of each sub-mm source at 0.6-arcsec resolution
are shown as contours. Following the assessment by Serjeant et al. of
the sub-mm signal-to-noise ratio of HDF850.3, we are not convinced by
its reality and therefore we do not include it.

We now discuss the individual sub-mm sources in, or close to the edge
of, the HDF as presented by Serjeant et al.\/ (2003). Our
discussion draws extensively on that of Serjeant et al.\/; however
in a number of cases we come to different conclusions as a result of
the improved radio information that we have at our disposal.

\subsubsection{HDF850.1}
\label{sect-HDF850.1}
The suggested association of Serjeant et al.\/ for this source is the
elliptical galaxy 3-586.0 (Williams et al.\/ 1998), although they do
allow the possibility that HDF850.1 is optically faint. The astrometric
alignment of the optical field to the ICRF (derived from radio
observations presented here), together with observations by IRAM
(again with respect to the ICRF) has now allowed Dunlop et al.\/
(2004) unambiguously to identify this brightest sub-mm source. Dunlop
et al.\/ have shown that, in a Subaru {\em K'}\, band image, from which a
smoothed version of WFPC2 {\em I}-band frame has been subtracted,
there is an extremely red object (ERO) with {\em I$-$K} $> 5.2$ at the
position of a very weak (16 $\mu$Jy) radio source which lies
$\sim1$~arcsec to the SW of the elliptical galaxy 3-586.0. The radio
source/ERO is within the 3$\sigma$ positional errors as measured by
the IRAM interferometer (Downes et al.\/ 1999) on the ICRF system. The
SED determined by Dunlop et al.\/ suggests the ERO is a starburst
system at a redshift of 4.1$\pm$0.5. The inferred star formation rate
is very high ($\sim$2000 M$_{\odot}$ yr$^{-1}$), although this may be reduced
by a factor close to 3 after correction for gravitational lens
boosting by the intervening elliptical galaxy $\sim$1~arcsec to the
NE.

\subsubsection{HDF850.2}
\label{sect-HDF850.2}
The SCUBA source lies just outside the HDF and hence we reproduce the
CFHT optical field; there is no optical candidate down to I$\sim$25
mag within the error circle. The nearest optical galaxy is 36564+1209
(I$\sim$23 mag, Barger et al.\/ 1999) which lies 5.6~arcsec to the
north of the SCUBA position. The extended, steep-spectrum radio source
J123656+621207 (46 $\mu$Jy, $\alpha>$1.32) lies 4.2~arcsec to the north
of the SCUBA position, outside the error circle. The radio source and
the optical galaxy are certainly not coincident.  It remains possible,
although in our view unlikely, that the radio source J123656+621207 is
associated with the sub-mm source. Whether the SCUBA source is
associated with J123656+621207 or not, it is clearly associated with
an optically faint system, below the limit of the CFHT or WFPC2 images
of this HFF field.

\subsubsection{HDF850.4}
\label{sect-HDF850.4}
HDF850.4 and HDF850.5 are a possible blended detection. Borys et al.\/
(2003) in a reworking of the sub-mm data list this as a single
detection (SMJ123650+621318). After consultation with S. Serjeant we
have treated HDF850.4 and HDF850.5 as separate sub-mm detections but
for completeness we have also plotted an error circle derived for the
blended source position from Borys et al.\/ (see above).

Serjeant et al.\/ (2003) suggest that HDF850.4 may be associated with
the $z=0.851$ (Cowie et al., \/2004) galaxy 2-339.0 which lies within
0.8~arcsec of the sub-mm source position and close to the centre of
the field as reproduced. However, Carilli \& Yun (1999) argue that
this is hard to reconcile with the lack of any detected radio emission
with an Arp 220-like SED since at this redshift they would expect
$S_{1.4}\sim280$ $\mu$Jy. There is a 3$\sigma$ radio detection of 14
$\mu$Jy close to the nuclear region of 2-339.0. However, this radio
flux density is still less than 1/20 of that predicted by Carilli
\& Yun (1999) for an Arp 220-type SED at $z=0.9$. We note that this
galaxy lies within the error circle associated with SMMJ123650+621318
(Borys et al.\/ 2003).

Another possible identification is with the bright {\em I}~=~21.1 mag galaxy
2-264.1 ($z=0.475$) coincident with the high signal-to-noise ratio
radio source J123649+621313 ($S_{1.4} = 49$ $\mu$Jy) which is listed as
a starburst system in Table~\ref{10arcminchars}. This galaxy lies
5.3~arcsec to the SW and within the error circle.

For completeness one should also consider the {\em I}~=~22.7 mag
($z=1.238$) galaxy which lies just outside the error circle
$\sim$6~arcsec to the west of the sub-mm position as a candidate
identification. This galaxy also contains a radio source ($S_{
1.4}\sim27$ $\mu$Jy). In addition, another candidate radio
identification lies close to the southern edge of the error
circle. The $S_{ 1.4}\sim19$ $\mu$Jy possible radio detection is not
associated with any optical object to the limit of the CFHT image.

The identification of this sub-mm source remains unclear.

\subsubsection{HDF850.5}
 \label{sect-HDF850.5}
Serjeant et al.\/ (2003) state that the position of this source is
likely to have been affected by the proximity of HDF850.4 (see also
above). The most obvious optical candidate, galaxy 2-404.0 ($z=0.199$,
Lanzetta et al.\/ 1996), lies 6.6~arcsec to the west but is
not a radio source.

Our deep radio image shows a possible radio counterpart on the eastern
edge of the error circle at the 3$\sigma$ level (14 $\mu$Jy). There is
no optical identification at this radio position to the limit of the
{\em HST} WFPC2 image.

We concur with Serjeant et al.\/ who conclude that there is no
reliable identification for this object.

\subsubsection{HDF850.6}
\label{sect-HDF850.6}
This source lies outside the HDF and we reproduce both the WFPC2 and
CFHT images of this region in the HFF. Serjeant et al.\/ (2003)
suggest that this object is most likely to be associated with the
radio source J123701+621146 ($S_{1.4}$~=~128 $\mu$Jy) which we
classify as a starburst in Table~\ref{10arcminchars}. The radio source
lies 2.9~arcsec from the sub-mm position but is within the error
circle. Although the source is extended over $\sim$3~arcsec in the
VLA-only image, a high resolution image shows that the central
1.5~arcsec overlies an ERO with {\em I--K}$>$5 (Alexander et al.\/
2001). This is not visible in the {\em HST} WFPC2 frame, but is just
detectable in the lower resolution CFHT image which has a better
surface brightness sensitivity. Cohen et al.\/ (2000) derive a
spectroscopic redshift of $z=0.884$ for the ERO on the basis of a
single detected emission line ({\sc o\,ii}, $\lambda_ {3727}$).

In agreement with Serjeant et al.\/ we conclude that J123701+621146
is the most likely radio counterpart to HDF850.6 and is probably
identified with the ERO and the SCUBA source. There is however one
other possible radio detection (15 $\mu$Jy) lying within the error
circle, but this has no optical counterpart to the limits of either of
the WFPC2 HFF or CFHT images. If this latter radio source is the
correct association, the radio to sub-mm flux ratio would imply a
redshift $z\geq$4 (see discussion in Section~\ref{sect-rad_submm}).

\subsubsection{HDF850.7}
 \label{sect-HDF850.7} Serjeant et al.\/ (2003) discuss a possible
 identification of this source with a $z=1.219$ (Barger et al.\/ 1999)
 {\em I}~=~22.3 mag galaxy which lies $\sim$5~arcsec from the sub-mm
 position just outside the error circle. This galaxy is identified
 with the steep-spectrum radio source J123634+621241 ($S_{1.4}=230$
 $\mu$Jy). Our high resolution image of this radio source confirms
 that it is an extended starburst system
 (Table~\ref{10arcminchars}). The galaxy is also detected in X-rays by
 {\em Chandra} and is identified with the {\em ISO} source HDF\_PM3\_3
 (Aussel et al.\/ 1999).

Although the positional offset gives a relatively high probability of
a chance coincidence with J123634+621241, Serjeant et al.\/ argue that
this source lies in a noisy area of the SCUBA image, and that the
positional errors may have been underestimated. Our radio image shows
only one other possible detection within the error circle (3$\sigma$,
15 $\mu$Jy) for which there is no optical counterpart to the limit of
the {\em HST} WFPC2 image. We conclude that the galaxy identified with
J123634+621241 is the most likely optical counterpart to HDF850.7, but
the positional offset makes this identification far from certain. By
analogy with HDF850.1, the weak radio detection cannot be discounted.

\subsubsection{HDF850.8}
\label{sect-HDF850.8}
Serjeant et al.\/ (2003) suggest that the galaxy 2-736.1, the
eastern-most member of an interacting pair of galaxies $\sim$2~arcsec
from the sub-mm position, is the most likely identification for this
source. Cohen et al.\/ (2000) measure a spectroscopic redshift of
$z=1.355$ for this galaxy.

Our radio image shows a weak ($\sim$15 $\mu$Jy) source associated with
2-736.1, although it is offset to the south-east by $\sim$0.5~arcsec
from the optical centroid. The radio to sub-mm flux ratio suggests a
higher redshift ($\sim$2) than that measured, although the
uncertainties are large. Although this remains the most plausible
identification for HDF850.8, there are a number of other possible
radio sources lying within the SCUBA error circle (including a 23-$\mu$Jy source on the western edge) which cannot be discounted. None,
however, have optical counterparts to the limit of the {\em HST} WFPC2
image.

\begin{table*}
\caption{Candidate radio source identifications above 4$\sigma$ in the
region of each sub-mm detection as listed by Serjeant et al.\/
(2003)$^{\bullet}$, Chapman et al. \/ (2004b)$^{\circ}$, and Borys et
al.\/ (2003)$^{\dagger}$. For Borys et al.\/ (2003) we include only
those sub-mm detections above 5$\sigma$ not already listed by either
Serjeant et al.\/ (2003) or Chapman et al.\/ (2004b). Furthermore, for
Borys et al.\/ (2003), since the nominal error circles are relatively
large, we include only those radio candidates above 5$\sigma$. Sources
marked with an * are possible bright (S$_{1.4}$>40 $\mu$Jy) radio
counterparts which lie just outside the nominal sub-mm positional
error circle as described in Section~\ref{sect-scuba} and shown in
Fig~\ref{scubapix}. The radio images were restored with a circular
beam of 0.6 arcsec prior to cross-referencing with the sub-mm
position. Redshift shown bracketed are uncertain due to significant
radio/sub-mm position offsets.}
\label{10arcminsubmm}
\begin{tabular}{cccccc}

&&&&&\\

{\bf Sub-mm Name}& S$_{850}$  & R.\/~A.\/  & Dec.\/ & Peak Radio Brightness & Redshift \\
                   & $\mu$Jy & \multicolumn{2}{c}{Radio}    &  $\mu$Jy beam$^{-1}$ & \\
\hline
&&&&&\\

 HDF850.1$^{\bullet}$ & 5.6 & 12:36:52.060\,\, & +62:12:25.67\,\,& 16 & 4.1\\
 HDF850.2$^{\bullet}$ & 3.5 & 12:36:56.5566    & +62:12:07.425 &   31*&\\
 HDF850.4$^{\bullet}$ & 1.1 & 12:36:49.455\,\, & +62:13:16.66\,\,& 20*& 1.238 \\
		      &     & 12:36:49.7432    & +62:13:13.065 &   27 & 0.475 \\
		      &     & 12:36:50.479\,\, & +62:13:16.10\,\,& 14 & 0.851 \\
		      &     & 12:36:50.647\,\, & +62:13:10.84\,\,& 19 \\
 HDF850.5$^{\bullet}$ & 1.0 & 12:36:52.852\,\, & +62:13:18.50\,\,& 14 \\
 HDF850.6$^{\bullet}$ & 6.4 & 12:37:01.021\,\, & +62:11:45.86\,\,& 14 \\
		      &     & 12:37:01.5745    & +62:11:46.814 &   34 \\
 HDF850.7$^{\bullet}$ & 5.5 & 12:36:34.5168    & +62:12:41.107 &  101 & 1.219 \\
		      &     & 12:36:34.812\,\, & +62:12:41.57\,\,& 15 \\
 HDF850.8$^{\bullet}$ & 1.7 & 12:36:52.393\,\, & +62:13:54.82\,\,& 23 \\
		      &     & 12:36:52.845\,\, & +62:13:54.00\,\,& 15 & 1.355 \\
		      &     & 12:36:52.907\,\, & +62:13:52.07\,\,& 13 \\
		      &     & 12:36:52.985\,\, & +62:13:57.97\,\,& 14 \\
		      &     & 12:36:53.123\,\, & +62:13:56.58\,\,& 14 \\ 
&&&&& \\ 
 SMMJ123606.9+621021$^{\circ}$& 11.6 & 12:36:06.8493 & +62:10:21.437 & 33 & 2.33 \\
 SMMJ123616.2+621514$^{\circ}$&  5.8 & 12:36:16.1419 & +62:15:13.937 & 32 & 3.61 \\
 SMMJ123618.3+621551$^{\circ}$&  7.3 & 12:36:18.3353 & +62:15:50.585 & 110 & (1.87) \\
 SMMJ123621.3+621708$^{\circ}$&  7.8 & 12:36:21.2691 & +62:17:08.458 & 116 & (1.99) \\
 SMMJ123622.7+621630$^{\circ}$&  7.7 & 12:36:22.653\,\,& +62:16:29.71\,\,& 25 & 2.40 \\
 SMMJ123629.1+621046$^{\circ}$&  5.0 & 12:36:29.124\,\,& +62:10:45.98\,\,& 23 & 1.013 \\
 SMMJ123635.6+621424$^{\circ}$&  5.5 & 12:36:35.5839 & +62:14:24.049 & 44 & 2.011 \\
 SMMJ123646.1+621449$^{\circ}$& 10.3 & 12:36:46.0629 & +62:14:48.713 & 91 & \\
 SMMJ123707.2+621408$^{\circ}$&  4.7 & 12:37:07.2209 & +62:14:08.208 & 31 & 2.48 \\
 SMMJ123712.0+621325$^{\circ}$&  4.2 & 12:37:11.9865 & +62:13:25.771 & 39 & 1.99 \\ 
&&&&& \\
 SMMJ123622+621618$^{\dagger}$ & 8.6 & 12:36:22.279\,\, & +62:16:15.57\,\, &   23 & \\
 SMMJ123634+621409$^{\dagger}$ & 11.2& \multicolumn{2}{c}{Not Detected} &$<$22 & \\
 SMMJ123637+621157$^{\dagger}$ & 7.0 & \multicolumn{2}{c}{Not Detected} &$<$22 & \\
 SMMJ123703+621303$^{\dagger}$ & 3.4 & 12:37:03.352\,\, & +62:13:06.15\,\, &   23 & \\
\hline
\end{tabular}

\noindent
\begin{flushleft}
\end{flushleft}
\end{table*}

\subsection{Other sub-mm samples}
\label{sect-other_submm_samples}

Chapman et al.\/ (2004c, submitted) have compiled a sample of sub-mm sources lying
within a 7.5-arcmin diameter field centred on R.\/~A.\/
12\h~36\m~48\fs0 Dec.\/ +62\degr~15\arcmin~40\farcs0 (J2000), which
have S$_{850}>$4 mJy and are detected at 1.4 GHz in the VLA A-array
observations of this region with S$_{1.4}>$40 $\mu$Jy (Richards
2000). This field centre lies 162 arcsec north and 10 arcsec east of
the HDF and 10-arcmin field pointing centre. Ten objects are common to
both the sub-mm sample and the 10-arcmin radio field.

In Table~\ref{10arcminsubmm} we list all the candidate radio
source/sub-mm identifications for Serjeant et al.\/ (2003) and Chapman
et al.\/ (2004b). For completeness we also note that Borys et al.\/
(2003) have reworked a large amount of sub-mm data in a field which
includes the HDF and parts of the HFF.  From their list of detections
we have selected those with a sub-mm signal-to-noise ratio $>$5 which
are not included in Serjeant et al.\/ (2003) and Chapman et al.\/
(2004b). These are listed in Table~\ref{10arcminsubmm} together with
any radio detections with a radio signal-to-noise ratio $>5$ falling
within an error circle of radius 1$\sigma$ obtained using figure 4 in
Borys et al.\/ (2003). These are the most plausible associations but
the large sub-mm positional errors listed by Borys et al.\/ (2003)
(typically $\sigma = 4 - 5$~arcsec) do not allow unambiguous
identifications to be made.

\subsection{The radio sub-mm connection}
\label{sect-rad_submm}
We have been able to combine the deepest 850-micron SCUBA images of
the HDF and HFF and the deepest, high resolution, 1.4-GHz radio
images, both with the best available astrometric accuracy. Despite
this only one sub-mm source, HDF850.1, can be unambiguously paired-up
with a radio source. The over-riding reason for this state-of-affairs
remains the present-day sub-mm resolution and astrometric accuracies
(with the exception of HDF850.1, separately measured with the IRAM
interferometer, Downes et al.\/ 1999) which are inadequate for crowded
fields. But this is not a complete explanation since in some cases
there are no plausible radio candidates within conservative sub-mm
error circles. Carilli \& Yun (1999, 2000) and Barger et al.\/ (2000)
have shown that the SEDs of both local and high-redshift ULIRGs are
similar, and that the ratio of the radio and sub-mm flux densities can
be used to construct a rough redshift indicator. The inverse {\em
K}-correction ensures that the sub-mm flux density varies little with
redshift, while the radio flux density falls off with the inverse
square law. Barger et al. in their figure 8 show that a galaxy with a
sub-mm flux density of 6 mJy will not be detected in the radio at 40
$\mu$Jy for a redshift $>$4.  Hence, it is likely that our failure to
detect radio counterparts for some of the sub-mm detections in the HDF
and HFF is simply due to their large redshift.

For the brighter sub-mm sources contained in the Chapman et al.\/
(2004b) sample, the positional errors are lower than for those of
Serjeant et al.\/ (2003) which has allowed them to link them with
radio sources in the 10-arcmin field. For the seven of their sub-mm
sources listed in Table~\ref{10arcminsubmm} with unambiguous
identifications, we find that the measured redshifts lie in range 1 to
4 with a mean of 2.26.

\section{OPTICALLY FAINT RADIO SOURCES}
\label{sect-faint}

13 of the 92 sources (14~per cent) in the 10-arcmin field are not
identified to the limit of the optical fields over which they
lie. Richards et al.\/ (1999) have suggested a number of possibilities
regarding the nature of the optically faint, unidentified $\mu$Jy
radio source population:

{\bf (i)} They are FRI or FRII radio galaxies at moderate
redshift. However, this would mean that optical counterparts would
have absolute magnitudes 6 magnitudes fainter than typical FRI or FRII
radio galaxies which have $I\sim$20 mag. Furthermore, our new 0.2-arcsec
resolution images do not have the appropriate radio morphology. We
therefore reject this hypothesis.

{\bf (ii)} They are displaced lobes of asymmetric sources. However,
other $\mu$Jy radio sources consist of only a single component
coincident with the optical counterpart, and the optically faint
systems do not have radio structures which differ significantly. We
therefore consider this hypothesis to be unlikely.

{\bf (iii)} The optical counterpart may be at such a high redshift,
$z>7$, that it is an {\em I}-band dropout due to the Gunn-Peterson effect.

{\bf (iv)} The optical counterpart may be obscured by dust -- see the
previous section on the radio  sub-mm connection.

{\bf (v)} They represent a new population of objects.\\

\noindent The last three possibilities remain to be tested. We can, however,
make some comments based on three individual objects.

As noted in Section~\ref{sect-z}, of the thirteen optically
faint systems in the 10-arcmin field, only two are classified as AGN
or AGN candidates. Six are classified as starbursts or starburst
candidates with the remaining five listed as unclassified. Subsequent
to the HDF WFPC2 imaging, deep IR studies of two of the starbursts
(J123642+621331 and J123651+621221) have revealed very red
objects and are now identified as high-redshift dust-obscured
starburst systems. In addition deep {\em K'}\, band imaging with the
Suburu 8-m telescope has revealed an extremely red object associated
with a weak 16-$\mu$Jy MERLIN+VLA 1.4-GHz radio detection at the
position of the brightest SCUBA sub-mm source HDF850.1 (see
Section~\ref{sect-HDF850.1} for a detailed description).

Waddington et al.\/ (1999) detect a very red object at the position of
J123642+621331 with NICMOS in {\em J}- and {\em H}-band and with the KPNO 4-m at
{\em K}-band. On the basis of a single Ly$\alpha$ emission line at 6595 \AA\/
and the observed spectral energy distribution, they infer a redshift
of 4.424, and argue that it is a distant starburst disc galaxy with a
weak embedded AGN. The radio structure consists of a compact component
together with emission extended over 0.4~arcsec and is thus
consistent with this interpretation. Garrett et al.\/ (2001) detect the
compact component with the EVN at an angular resolution of 20 mas;
this may be the steep-spectrum AGN core.

Dickinson (2000) finds a very red NICMOS detection at the position of
J123651+621221. The spectral energy distribution implies a
dust-shrouded starburst at a redshift close to 2.7 (Alexander et al.\/
2001).  The source is a hard {\em Chandra} X-ray detection which
Hornschemeier et al.\/ (2001) interpret as an obscured QSO. However,
the extended steep-spectrum radio structure suggests a dust-obscured
starburst system with no evidence for any compact radio core. If there
is an obscured QSO in this object which is responsible for the X-ray
emission, it is not a significant radio source.

The nature of the remaining optically faint sources remains unclear,
but from the radio structures seen in the MERLIN+VLA combination
images, it seems likely that those classified as starburst candidates
are most likely to be high-redshift dusty systems. These represent the
first few detections of a new population of objects. A variety of
explanations are possible, for example activity around young black
holes at $z>10$, perhaps resulting from Population III star clusters.
Following Haiman, Dijkstra \& Mesinger (2004), such objects could be
detectable at a level of tens of $\mu$Jy at a high enough surface
density to occur in these observations.  Such possibilities underscore
the vital role played by high resolution radio observations since they
have high astrometric positional accuracy, show the detailed
distribution of galactic radio activity, and do not suffer from dust
obscuration. However, more sensitive observations, with higher angular
resolution, in the IR and sub-millimetre bands are required to
understand the true nature of these optically faint radio sources.

\section{FUTURE DEVELOPMENTS}
\label{sect-future-radio-obs}

Using the same data and with the availability of enhanced computer
power, we are extending the statistical study limited to the 3-arcmin
field in this paper (see Section~\ref{sect-3arcmin} and
Fig.~\ref{fig5}), to encompass the whole of the 10-arcmin field. This
will form the basis of a future paper and will contain a statistical
analysis of a mapped area over an order of magnitude larger than the
3-arcmin field. Very recently deeper {\em HST} ACS images have been
released (Giavalisco et al 2004), which we will compare with the
radio data.  We do not wish to delay the present
publication, since the new {\em HST} data require astrometric
alignment in order to have confidence in any further identifications
of radio sources with faint optical galaxies.

With the development of ${\it e}$-MERLIN and the EVLA significantly
more sensitive observations of the HDF and HFF will become feasible
within a few years. When these broad-band instruments come on line,
imaging at 1.4 GHz will be a factor of five times more sensitive than
now. Combination imaging with similar integration times to the present
study will therefore yield rms noise levels $\sigma_{e\rm
-MERLIN+EVLA}\sim$0.6 $\mu$Jy beam$^{-1}$ in which radio sources like
that associated with HDF 850.1, with a flux density of only 16
$\mu$Jy, will be detected at the 20$\sigma$ level. The slope of the
integral source counts below 40 $\mu$Jy implies an increase in the
number of radio sources by a factor of six in a sample complete to 8
$\mu$Jy (equivalent to the 40-$\mu$Jy selection limit in the present
10-arcmin field). To this limit in the 10-arcmin field, we will be
able to image in excess of 500 radio sources. Above 5 $\mu$Jy
(8$\sigma$ -- equivalent to the 27-$\mu$Jy limit in the 3-arcmin
field) we will image around 1000 radio sources. Furthermore, the
radio-optical statistical studies will be extended to radio sources
with flux densities less than 1 $\mu$Jy.

Compared with our current capabilities, at 5 GHz the imaging
sensitivity will be increased by a factor of $\sim$30 within the inner
3-arcmin field (constrained by the Lovell telescope primary beam), and
by a factor of $\sim$15 within a 9-arcmin field (constrained by the
primary beams of the other antennas). This will result in noise levels
of $\sigma_{e\rm -MERLIN+EVLA}\sim$0.15 $\mu$Jy beam$^{-1}$ in the inner
region and 0.3 $\mu$Jy beam$^{-1}$ in the outer combined with an angular
resolution of $\sim$50 mas. Many of the starburst systems resolved for
the first time in this work will thus be mapped with very high angular
resolution, revealing the detailed areas of star-formation for the
first time and detecting any embedded AGN compact cores. Together with
the 5-GHz astrometric precision of a few mas, detailed cross-waveband
comparisons of conditions within each galaxy will thus be possible.

\section{SUMMARY AND CONCLUSIONS}
\label{sect-concl}

\subsection{Observational Results}
\label{sect-observational_results_summary}

In a 10-arcmin field centred on the HDF, a complete sample of 92
sources with flux densities above 40 $\mu$Jy have been detected with
the VLA, and imaged with the MERLIN+VLA combination at resolutions of
0.2, 0.3 and 0.5 arcsec (Table~\ref{10arcmin.tab}). These are amongst
the most sensitive 1.4-GHz images yet made, with rms noise levels of
3.3 $\mu$Jy beam$^{-1}$ in the 0.2-arcsec resolution images. The
images also combine high resolution (down to 0.2~arcsec) with high
astrometric accuracy ($<50$ mas in the HDF) and allow us to probe the
association of radio source types with both the optical galaxy
population and the sub-mm source population within the HDF and HFF. We
note that a comparison of our catalogue with one from WSRT at lower
angular resolution suggests that we may be missing $\sim$10 per cent of the
sources which have low radio surface brightness and angular sizes in
the range 5 -- 20 arcsec.

Positions derived from the independent MERLIN and VLA imaging agree to
better than 15 mas over the entire 10-arcmin field. Radio sources
associated with compact galaxies have been used to align the WFPC2 and
CFHT optical fields to the ICRF (Fig.~\ref{fig3} and
Table~\ref{10arcminerrs}). The optical field has been aligned to
better than 50 mas in the Deep Field itself, and to $\sim$150 mas in
the outer parts of the Flanking Fields. The combined optical and radio
images are reproduced in Fig.~\ref{colourpix}.

\subsection{Interpretation}
\label{sect-interpretation_summary}

\noindent {\it\textbf{Radio source characteristics}}:\\
The available radio spectral and morphological data have enabled us to
classify 72 per cent of the sources in the 10-arcmin field as either definite
or candidate AGNs (20 per cent), or starburst systems (52 per cent). The radio
source sizes are typically $\sim$1~arcsec (Fig.~\ref{fig4}), somewhat
smaller than the optical galaxy images. In addition to the imaging of
92 sources found with the VLA alone, the central 3 arcmin, which
includes the HDF, has been separately imaged with the MERLIN+VLA
combination at the full 3.3 $\mu$Jy beam$^{-1}$ sensitivity to search
for sources down to 27 $\mu$Jy. Although an additional $\sim$8 sources
are expected to be present from number counts, no additional sources
were found that were not previously detected by the VLA, indicating
that such sources are heavily resolved with MERLIN and hence must have
angular sizes $\ga1$~arcsec.

The typical rest-frame luminosity of the AGN and AGN candidate sources
is similar to FRI type systems, but is significantly less than that of
powerful FRII type radio galaxies and radio-loud quasars. Only two AGN
systems show `classical' double radio structures; the vast majority
being small core-jet structures extended on sub-galactic scales. All
the starburst and starburst candidate sources are substantially more
luminous than the nearby starburst galaxy M82, and about half are
more luminous than the powerful ULIRG Arp 220. A number of starburst
systems at redshifts in excess of 2 have been identified (Chapman et
al.\/ 2004a, 2004c submitted) as sub-mm sources with the aid of these
radio data and are several orders of magnitude more luminous than Arp
220.

The proportion of starburst systems is found to increase with
decreasing source strength (Fig.~\ref{sourceclass}). At flux densities
below 100 $\mu$Jy more than 70~per cent of the $\mu$Jy sources are
starburst-type systems associated with major disc galaxies in the
redshift range 0.3--1.3. Some 40~per cent of the brighter sources are
found to be intermediate luminosity AGN systems identified with
galaxies in a similar redshift range.

\vspace*{5mm}

\noindent {\it\textbf{Optical identifications}}:\\
We find a {\it statistical} association of very faint ($\ge$2 $\mu$Jy)
radio sources with optically brighter HDF galaxies down to $\sim$23
mag (Fig.~\ref{fig5}). Of the 92 sources in the 10-arcmin field above
40 $\mu$Jy, around 85~per cent are identified with galaxies brighter
than {\em I}~=~25 mag. The remaining 15~per cent are associated with
optically faint systems close to or beyond the HFF limit (and for some
objects, the HDF limit). As discussed in Section~\ref{sect-chandra}
many of these may be dust-shrouded starburst galaxies at high redshift
(z$>$3). The high astrometric accuracy of the alignment and the
ability of radio to see through obscuring dust have proved to be vital
in making the correct optical identification of these very faint radio
objects. In particular, this has led to the correct identification of
three optically faint, very red systems (J123642+621331,
J123651+621221, and SCUBA HDF850.1). This underscores the vital role
to be played by radio observations in probing the distant galaxy
population.

\vspace*{5mm}

\noindent {\it\textbf{X-ray identifications}}:\\
Over half of the radio sources in the 10-arcmin field are detected
in X-rays by {\em Chandra}, however the X-ray detection rate appears
to be uncorrelated with the radio source classification or flux
density. This does not support the hypothesis that an X-ray detection
is a direct diagnostic of AGN activity unless X-ray and radio emission
mechanisms have different origins within the same galaxy.  This will
be investigated further in a later paper.

\vspace*{5mm}

\noindent {\it\textbf{Sub-mm identifications}}:\\
We have combined our aligned radio and optical fields with the latest
sub-mm results on the HDF and HFF. Currently radio observations can
provide an important indicator of the correct identification of a high
redshift galaxy (eg SCUBA HDF850.1, Dunlop et al. 2004). However, even
with the aid of these ultra-sensitive radio observations and excellent
radio-optical astrometry, reliable identifications of faint sub-mm
sources in the HDF and HFF remain elusive. This will remain the case
until sub-arcsecond astrometry in this waveband becomes routinely
available with the advent of ALMA.

\section*{Acknowledgements}
This paper has made use of two radio arrays: MERLIN, a UK National
Facility operated by the University of Manchester at Jodrell Bank
Observatory on behalf of PPARC, and the VLA of the National Radio
Astronomy Observatory a facility of the National Science Foundation
operated under cooperative agreement by Associated Universities, Inc.
The optical data are based in part on observations made with the
NASA/ESA Hubble Space Telescope, obtained at the Space Telescope
Science Institute, which is operated by the Association of
Universities for Research in Astronomy, Inc., under NASA contract
NAS5-26555; and in part on observations made with the
Canada-France-Hawaii Telescope by Barger et al.\/ (1999).  The
cross-matching of the radio source list with the {\em ISO} (Aussel et
al.\/ 1999) and {\em Chandra} (Alexander et al. 2003) catalogues and other
multi-wavelength comparisons made use of software tools developed by
the Astronomical Virtual Observatory (www/euro-vo.org).  We thank the anonymous referee for helpful suggestions.

\bibliographystyle{mnras}

\clearpage

\onecolumn
\appendix
\section{Properties of Radio Sources in the 10-arcmin Field}
\begin{table*}
\centering
\caption{{\bf Radio Sources in the 10-arcmin Field: Astrometric
Characteristics}. The MERLIN+VLA peak position is given unless the
source is a VLA-only detection (marked {\em V}); $\sigma_{\rm pos}$ is
the error in the position in R.\/~A.\/ and Dec.\/ Beam is
the FWHM of the circular restoring beam used for the radio images in
Fig.~\ref{colourpix}. `Optical' refers to the telescope providing the optical image, either
C for CFHT (Barger et al.\/ 1999) or H for {\em HST} WFPC2 (Williams et
al. 1996).}
\label{10arcmin.tab}
\begin{tabular}{lllrcc}
&&&&&\\
Name  & \multicolumn{2}{l}{MERLIN+VLA Position (Peak)} & $\sigma_{\rm pos}$ & Beam & Optical \\
      & R.\/A. & Dec.\/ &(mas) & (\arcsec )&  \\
\hline &&&&&\\
J123606+620951 &12 36 06.6128 & +62 09 51.141 &    4,\,\,\,\,\,\, 5 & 0.2 & -- \\
J123606+621021 &12 36 06.8493 & +62 10 21.437 &   39,\,\,\,   29    & 0.5 & -- \\
J123607+621328 &12 36 07.1427 & +62 13 28.632 &   34,\,\,\,  62     & 0.5 & -- \\
J123608+621035 &12 36 08.1195 & +62 10 35.898 &    4,\,\,\,\,\,\, 4 & 0.3 & C  \\
J123608+621553 &12 36 08.2421 & +62 15 53.094 &   47,\,\,\,   45    & 0.5 & C  \\
&&&&&\\												    
J123608+621431 &12 36 08.9413 & +62 14 31.026 &   78,\,\,\,   38    & 0.5 & C  \\
J123610+620810 &12 36 10.5718 & +62 08 10.726 &   14,\,\,\,  15     & 0.3 & -- \\
J123610+621651 &12 36 10.5514 & +62 16 51.669 &   38,\,\,\,   30    & 0.5 & C  \\
J123612+621138 &12 36 12.0272 & +62 11 38.733 &   80, 108           & 0.5 & C  \\
J123612+621140 &12 36 12.4884 & +62 11 40.487 &   49,\,\,\,  47     & 0.5 & C  \\
&&&&&\\												    
J123615+620946 &12 36 15.6276 & +62 09 46.803 &   71,\,\,\,  53     & 0.5 & C  \\
J123616+621513 &12 36 16.1419 & +62 15 13.937 &   51,\,\,\,  53     & 0.5 & C  \\
J123617+621011 &12 36 17.0801 & +62 10 11.306 &   32,\,\,\,  28     & 0.5 & C  \\
J123617+621540 &12 36 17.5541 & +62 15 40.768 &    5,\,\,\,\,\,\, 5 & 0.5 & C  \\
J123618+621635 &12 36 18.017{\em V} & +62 16 35.27{\em V}&56,\,\,\, 41&0.5 & C  \\
&&&&&\\												    
J123618+621550 &12 36 18.3353 & +62 15 50.585 &    4,\,\,\,\,\,\, 4 & 0.2 & C  \\  
J123619+621252 &12 36 19.4784 & +62 12 52.581 &   13,\,\,\,  11     & 0.2 & C  \\
J123620+620844 &12 36 20.2622 & +62 08 44.250 &    3,\,\,\,\,\,\, 3 & 0.2 & C  \\
J123621+621109 &12 36 21.2256 & +62 11 09.007 &   29,\,\,\,  27     & 0.5 & C  \\
J123621+621708 &12 36 21.2691 & +62 17 08.458 &    5,\,\,\,\,\,\, 5 & 0.3 & C  \\ 
&&&&&\\												    
J123622+621544 &12 36 22.4753 & +62 15 44.776 &   41,\,\,\,  49     & 0.5 & H  \\
J123622+621629 &12 36 22.6535 & +62 16 29.718 &   78,\,\,\,  69     & 0.5 & C  \\
J123622+620945 &12 36 22.7788 & +62 09 45.756 &   63,\,\,\,  53     & 0.5 & C  \\
J123623+621642 &12 36 23.5436 & +62 16 42.754 &    1,\,\,\,\,\,\, 1 & 0.3 & C  \\
J123624+621017 &12 36 24.2913 & +62 10 17.262 &   53,\,\,\,  38     & 0.5 & H  \\
&&&&&\\												    
J123624+621743 &12 36 24.7685 & +62 17 43.160 &   22,\,\,\,  27     & 0.5 & C  \\
J123629+621045 &12 36 29.1240 & +62 10 45.984 &    7,\,\,\,  50     & 0.5 & C  \\
J123630+620923 &12 36 30.0516 & +62 09 23.895 &   46,\,\,\,  52     & 0.5 & C  \\
J123630+620851 &12 36 30.4861 & +62 08 51.051 &   69,\,\,\,  62     & 0.5 & C  \\
J123631+620957 &12 36 31.2450 & +62 09 57.791 &    8,\,\,\,  10     & 0.2 & H  \\
&&&&&\\												    
J123632+621658 &12 36 32.412{\em V} & +62 16 58.59{\em V}&  140,\,\,140&0.5 & C  \\
J123632+620759 &12 36 32.5583 & +62 07 59.846 &   18,\,\,\,  25     & 0.3 & -- \\
J123633+621005 &12 36 33.7269 & +62 10 05.962 &   67,\,\,\,  63     & 0.5 & H  \\
J123634+621213 &12 36 34.4701 & +62 12 13.006 &   16,\,\,\,  22     & 0.3 & H  \\
J123634+621241 &12 36 34.5168 & +62 12 41.107 &   11,\,\,\,  11     & 0.2 & H  \\
&&&&&\\												    
J123635+621424 &12 36 35.5839 & +62 14 24.049 &    7,\,\,\,\,\,\, 7 & 0.2 & H  \\
J123636+621320 &12 36 36.9061 & +62 13 20.337 &   35,\,\,\,  28     & 0.5 & H  \\  
J123637+620852 &12 36 36.9973 & +62 08 52.417 &   36,\,\,\,  26     & 0.5 & C  \\
J123640+621009 &12 36 40.6888 & +62 10 09.909 &   43,\,\,\,  40     & 0.5 & H  \\
J123641+620948 &12 36 41.5511 & +62 09 48.232 &   14,\,\,\,  16     & 0.3 & H  \\

\end{tabular}
\end{table*}

\begin{table*}
\centering
\contcaption{\bf Radio Sources in the 10-arcmin Field: Astrometric
Characteristics.}
\begin{tabular}{lllrcc}

&&&&&\\

Name  & \multicolumn{2}{l}{MERLIN+VLA Position (Peak)} & $\sigma_{\rm pos}$ & Beam & Optical \\
      & R.\/A. & Dec. &(mas) & (\arcsec )&  \\

\hline &&&&&\\

J123642+621331 &12 36 42.0916 & +62 13 31.426 &    1,\,\,\,\,\,\, 1 & 0.2 & H  \\   
J123642+621545 &12 36 42.2123 & +62 15 45.521 &   11,\,\,\,  12     & 0.5 & C  \\
J123644+621133 &12 36 44.3870 & +62 11 33.145 &    1,\,\,\,\,\,\, 1 & 0.2 & H  \\ 
J123645+620754 &12 36 45.862{\em V} & +62 07 54.19{\em V}&89,\,\,\, 68&0.5 & -- \\
J123646+621448 &12 36 46.0629 & +62 14 48.713 &    5,\,\,\,\,\,\, 4 & 0.2 & C  \\
&&&&&\\												    
J123646+621629 &12 36 46.3344 & +62 16 29.374 &   19,\,\,\,  19     & 0.5 & C  \\ 
J123646+621404 &12 36 46.3321 & +62 14 04.693 &    2,\,\,\,\,\,\, 2 & 0.2 & H  \\
J123646+620833 &12 36 46.6587 & +62 08 33.291 &   49,\,\,\,  49     & 0.5 & C  \\     
J123646+621226 &12 36 46.673{\em V} & +62 12 26.28{\em V}&  190,\,\,190&2.0& H  \\
J123646+621445 &12 36 46.736{\em V} & +62 14 45.64{\em V}&70,\,\,\, 68&0.5 & H  \\ 
&&&&&\\
J123649+621313 &12 36 49.7432 & +62 13 13.065 &   42,\,\,\,  30     & 0.5 & H  \\
J123650+620801 &12 36 50.1335 & +62 08 01.973 &   19,\,\,\,  19     & 0.5 & -- \\
J123650+620844 &12 36 50.1886 & +62 08 44.601 &   37,\,\,\,  35     & 0.3 & C  \\
J123651+621030 &12 36 51.1223 & +62 10 30.955 &   29,\,\,\,  32     & 0.5 & H  \\
J123651+621221 &12 36 51.7258 & +62 12 21.435 &   34,\,\,\,  18     & 0.5 & H  \\
&&&&&\\												   
J123652+621444 &12 36 52.8839 & +62 14 44.076 &    6,\,\,\,\,\,\, 6 & 0.5 & H  \\
J123653+621139 &12 36 53.3629 & +62 11 39.647 &   14,\,\,\,  10     & 0.2 & H  \\
J123654+621040 &12 36 54.6828 & +62 10 40.429 &   38,\,\,\,  53     & 0.5 & H  \\
J123655+620917 &12 36 55.7457 & +62 09 17.478 &   16,\,\,\,  23     & 0.3 & C  \\
J123655+620808 &12 36 55.9397 & +62 08 08.163 &   10,\,\,\,  13     & 0.3 & -- \\
&&&&&\\												   
J123656+621207 &12 36 56.5566 & +62 12 07.425 &   41,\,\,\,  34     & 0.5 & H  \\
J123656+621301 &12 36 56.9170 & +62 13 01.783 &   97,\,\,\,  75     & 1.0 & H  \\
J123659+621449 &12 36 59.9150 & +62 14 49.503 &   68,\,\,\,  86     & 0.5 & H  \\
J123700+620909 &12 37 00.2480 & +62 09 09.778 &    1,\,\,\,\,\,\, 1 & 0.3 & C  \\
J123701+621146 &12 37 01.5745 & +62 11 46.814 &   38,\,\,\,  78     & 0.5 & H  \\
&&&&&\\												   
J123702+621401 &12 37 02.759{\em V} & +62 14 01.63{\em V}&40,\,\,\, 40&0.5 & H  \\
J123704+620755 &12 37 04.1120 & +62 07 55.484 &   64,\,\,\,  91     & 0.5 & -- \\
J123705+621153 &12 37 05.8599 & +62 11 53.541 &   79,\,\,\,  83     & 0.5 & H  \\
J123707+621408 &12 37 07.2209 & +62 14 08.208 &   60,\,\,\,  49     & 0.5 & C  \\
J123707+621121 &12 37 07.9867 & +62 11 21.743 &   28,\,\,\,  33     & 0.3 & C  \\
&&&&&\\												   
J123708+621056 &12 37 08.3663 & +62 10 56.049 &   20,\,\,\,  21     & 0.5 & H  \\
J123709+620837 &12 37 09.4315 & +62 08 37.580 &    6,\,\,\,\,\,\, 6 & 0.3 & C  \\
J123709+620841 &12 37 09.7518 & +62 08 41.249 &   16,\,\,\,  11     & 0.3 & C  \\
J123711+621330 &12 37 11.2549 & +62 13 30.846 &   12,\,\,\,  14     & 0.3 & C  \\
J123711+621325 &12 37 11.9865 & +62 13 25.771 &   32,\,\,\,  38     & 0.5 & C  \\
&&&&&\\												   
J123713+621603 &12 37 13.5934 & +62 16 03.689 &   90,\,\,\,  87     & 0.5 & C  \\
J123714+621558 &12 37 14.3371 & +62 15 58.846 &   52,\,\,\,  77     & 0.5 & C  \\
J123714+620823 &12 37 14.9414 & +62 08 23.208 &    1,\,\,\,\,\,\, 1 & 0.2 & C  \\
J123716+621512 &12 37 16.3740 & +62 15 12.343 &    2,\,\,\,\,\,\, 2 & 0.3 & C  \\
J123716+621643 &12 37 16.5410 & +62 16 43.799 &   31,\,\,\,  51     & 0.5 & C  \\
&&&&&\\												   
J123716+621733 &12 37 16.6811 & +62 17 33.327 &    2,\,\,\,\,\,\, 2     & 0.3 & C  \\
J123716+621007 &12 37 16.8252 & +62 10 07.401 &   19,\,\,\,  22     & 0.5 & C  \\
J123717+620827 &12 37 17.520{\em V} & +62 08 27.61{\em V}&51,\,\,\, 91&0.5 & C  \\
J123719+620902 &12 37 19.9643 & +62 09 02.736 &   62,\,\,\,  43     & 0.5 & C  \\
J123721+621129 &12 37 21.2539 & +62 11 29.954 &   1,\,\,\,\,\,\,  1 & 0.3 & H  \\
&&&&&\\												   
J123721+621346 &12 37 21.4669 & +62 13 46.675 &   58,\,\,\,  54     & 0.5 & C  \\
J123725+620856 &12 37 25.0112 & +62 08 56.374 &   24,\,\,\,  26     & 0.3 & C  \\
J123725+621005 &12 37 25.298{\em V} & +62 10 05.91{\em V}&98,\,\,\, 60&0.5 & C  \\
J123725+621128 &12 37 25.9495 & +62 11 28.705 &    2,\,\,\,\,\,\, 2 & 0.2 & C  \\
J123730+621258 &12 37 30.7907 & +62 12 58.804 &    4,\,\,\,\,\,\, 5 & 0.2 & -- \\
&&&&&\\												   
J123732+621012 &12 37 32.644{\em V} & +62 10 12.84{\em V}&  180,\,\,\,180& 0.5 &-- \\
J123734+620931 &12 37 34.242{\em V} & +62 09 31.93{\em V}&  140,\,\,\,140& 0.5 &-- \\
&&&&&\\												  
													  
\end{tabular}												  

\end{table*}

\newpage

\begin{table*}
\vspace*{-1.7cm}
\resizebox{19.1cm}{!}{
\epsfbox{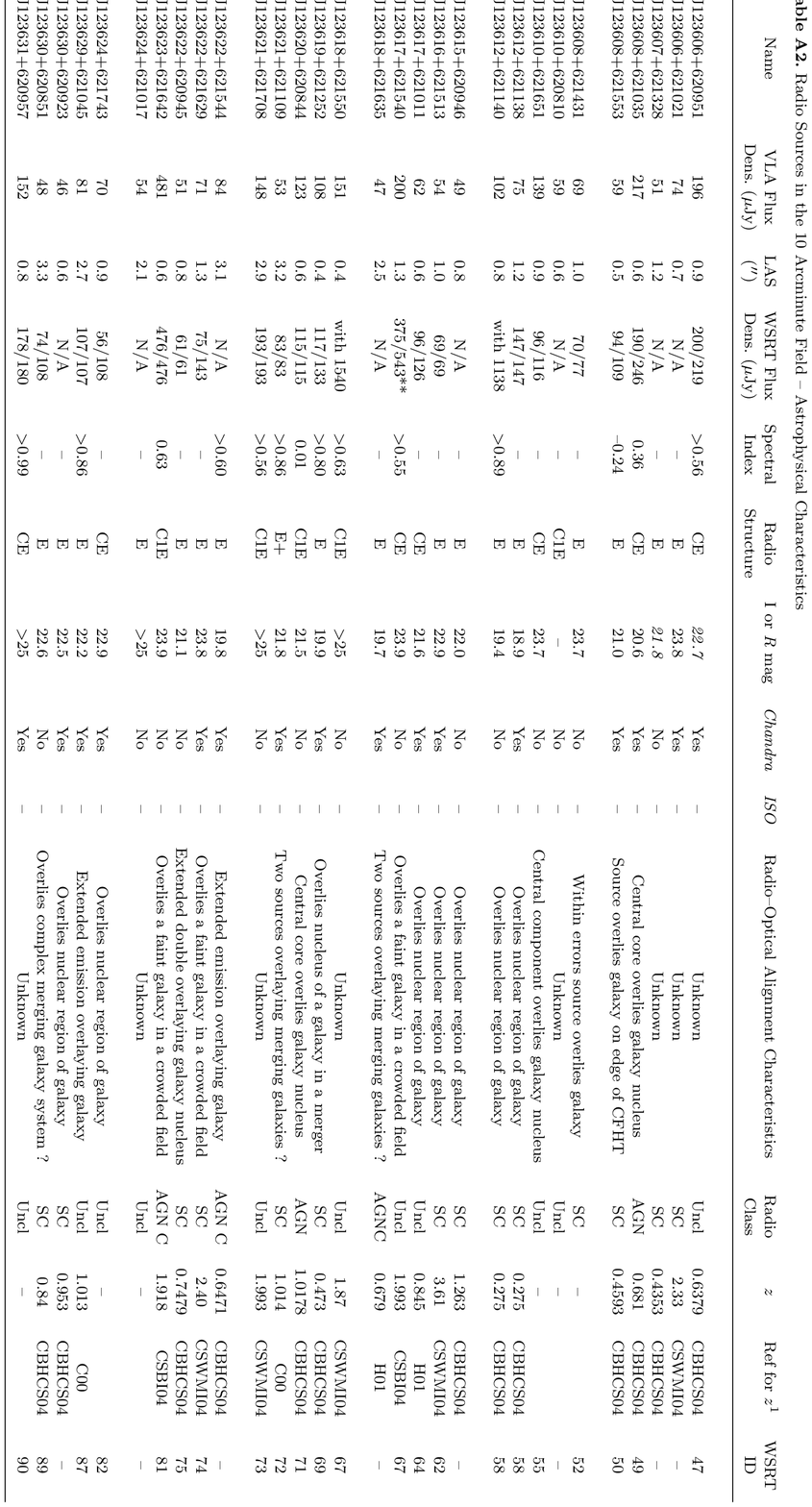}}
\vspace*{1.5cm}
\caption{}
\label{10arcminchars}
\end{table*}
\newpage

\begin{table*}
\vspace*{-1.7cm}
\resizebox{19.1cm}{!}{
\epsfbox{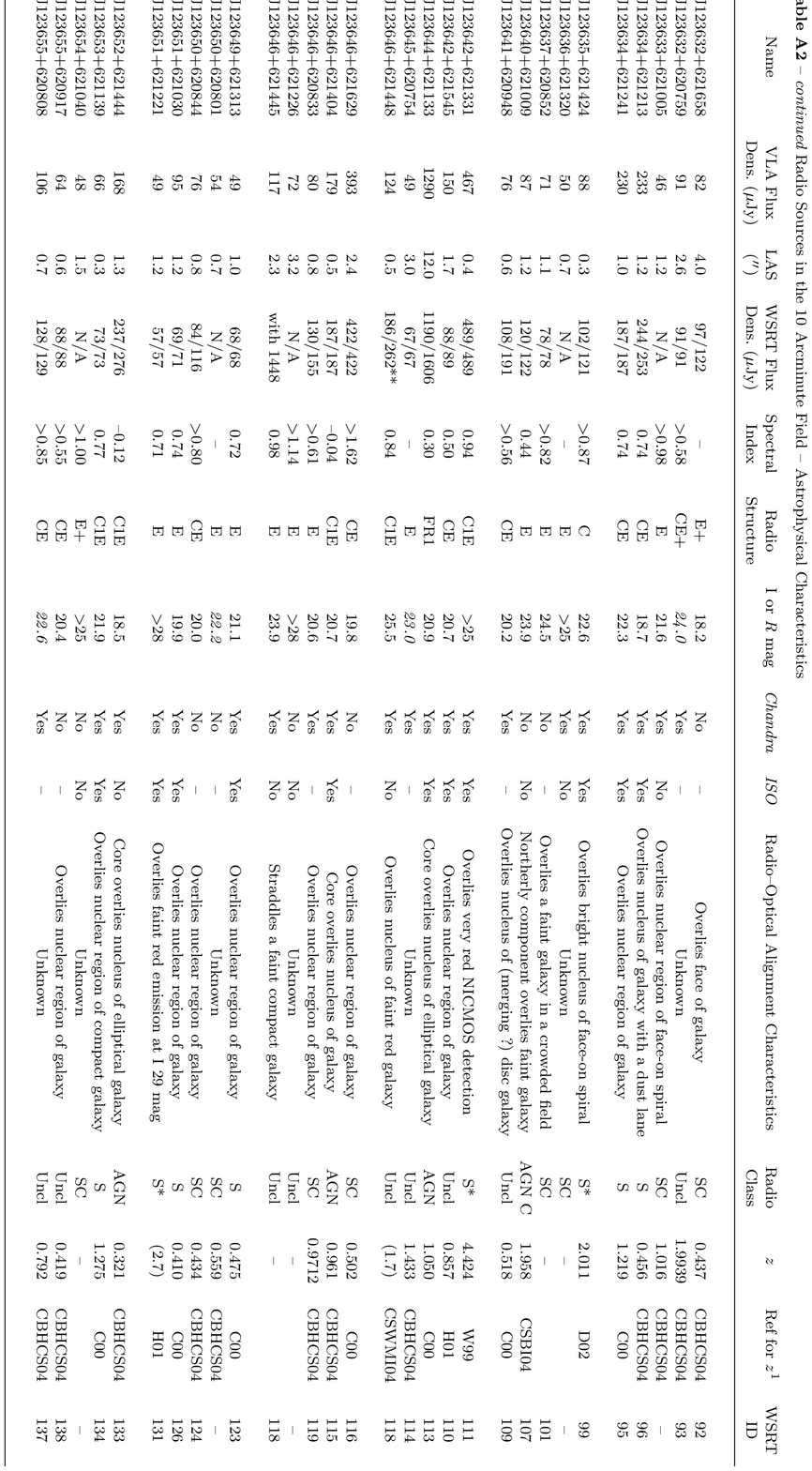}}
\end{table*}
\newpage

\begin{table*}
\vspace*{-1.7cm}
\resizebox{19.1cm}{!}{
\epsfbox{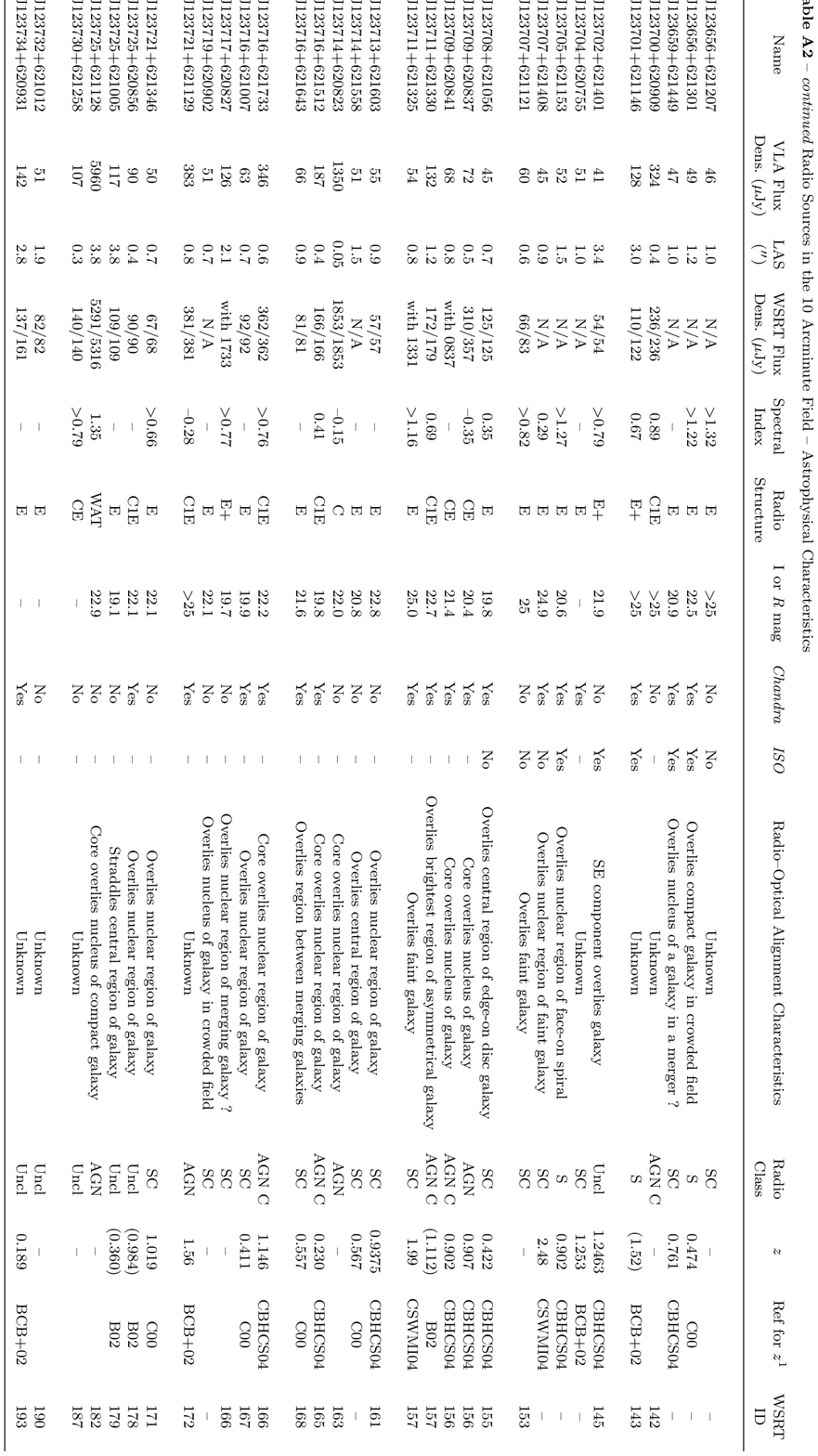}}
\end{table*}
\setcounter{table}{2}

\begin{table}
\contcaption{Radio Sources in the 10 Arcminute Field -- Astrophysical Characteristics:}
$^1$References for redshifts $z$\\
\\
\begin{tabular}{ll}
BCB+02& Barger et al.\/ (2002)\\
B02 &Bauer et al.\/ (2002)\\
CSBI04 &Chapman et al.\/(2004a) (received after submission; data not included in figures.)\\
CSWMI04& Chapman et al.\/(2004b) (received after submission; data not included in figures.)\\
C00 &Cohen et al.\/ (2000)\\
CBHCS04& Cowie et al.\/ (2004); Team Keck Redshift Survey {\tt http://www2.keck.hawaii.edu/realpublic/science/tksurvey/index.html}\\
D02 &Dawson et al.\/ (2002)\\
H01 &Hornschemeier et al.\/ (2001)\\
W99 &Waddington et al.\/ (1999)\\
\end{tabular}
\end{table}

\clearpage
\newpage
\twocolumn
\section{Source details for the 10-arcminute field}
\label{sect-details}
In this section we give a brief description of the properties of each
source detected and imaged in the 10-arcmin field. Where known, radio
spectral indices are given ($S\propto\nu{^-{^\alpha}}$). For weaker sources
close to the edge of the 10-arcmin field, spectral index information
is unavailable since they are on the edge of the VLA primary beam
response at 8.4 GHz and are thus undetected by Richards et al.\/
(1998). For stronger sources in this region, which still remain
undetected at 8.4 GHz, lower limits to the radio spectral index are
given.  We note other detections and compare the radio and optical
morphologies; however we cannot assume that the emission at different
wavelengths comes from the same part of the galaxy or the same
mechanism unless there is additional evidence for this, including the
tight radio-far-IR correlation (see references in
Section~\ref{sect-iso}).  Table~\ref{10arcminchars} gives additional
information (such as whether non-detections at other wavelengths are
due to the limited fields of view of the relevant instruments). {\em
ISO} data are taken from Aussel et al.\/ (1999) and {\em Chandra} data
are taken from Hornschemeier et al.\/ (2001) and Alexander et al.\/
(2003).  We have not yet included the latest {\em HST} ACS data for
reasons noted in Section~\ref{sect-future-radio-obs}.

{\bf J123606+620951} (Unclassified): is undetected at 8.4 GHz and
thus has a radio spectrum with $\alpha>0.56$. It has a compact ($<$0.2~arcsec)
component with weak extended emission. The radio source
lies outside the CFHT image so no optical frame is shown in
Fig.~\ref{colourpix}. However Cowie et al.\/ (2004) catalogue an {\em
R} = 22.7 mag galaxy at the position of the radio source with a
redshift of 0.6379. It is detected by {\em Chandra}.

{\bf J123606+621021} (Starburst Candidate): is separated from
J123608+621035 by 17~arcsec but does not appear to be related to
it. The radio structure is extended over 0.7 arcsec. It lies outside
the area of the CFHT image so no optical frame is shown in
Fig.~\ref{colourpix}. However Chapman et al.\/ (2004b) catalogue an
I=23.8 mag sub-mm galaxy (S$_{850}=11.6\pm3.5$ mJy) at the position of
the radio source with a redshift of 2.33.  There is a faint {\em
Chandra} counterpart.

{\bf J123607+621328} (Starburst Candidate): is extended over
1.2~arcsec roughly NS. This source  lies outside the CFHT image so
no optical frame is shown in Fig.~\ref{colourpix}. However Cowie et
al.\/ (2004) catalogue an {\em R} = 21.8 mag galaxy at the position of the
radio source with a redshift of 0.4353.

{\bf J123608+621035} (AGN): has a relatively flat radio spectrum
overall ($\alpha$=0.36).  It has a compact core and two-sided radio
emission.  The compact component lies at the nucleus of a 20.6 mag
galaxy with a redshift of 0.681 (Cowie et al.\/ 2004).  It is detected
by {\em Chandra}.

{\bf J123608+621553} (Starburst Candidate): Richards (2000) states
that this source has an inverted radio spectrum ($\alpha=-0.24$).  It
is extended by approximately 0.5~arcsec.  It is identified with a {\em
I}~=~21 mag galaxy at the very edge of the CFHT frame with a redshift
of 0.4593 (Cowie et al.\/ 2004). No compact AGN core is seen,
therefore it is most likely that the radio emission originates in an
extended starburst with strong free-free absorption at 1.4 GHz to
account for the spectral shape.  It is detected by {\em Chandra}.

{\bf J123608+621431} (Starburst Candidate): is extended by
$\sim$1~arcsec EW and is identified with a 23.7 mag galaxy of the CFHT
frame.

{\bf J123610+620810} (Unclassified): has a relatively compact
component with weak extended emission $\sim$0.6~arcsec to the SW
at 0.5-arcsec resolution (not shown). No optical information is
available since it lies outside the area of the CFHT frame.

{\bf J123610+621651} (Unclassified): has extended radio emission
(0.9~arcsec) on both sides of a relatively compact component at
0.2-arcsec resolution (not shown), which is coincident with the
nuclear region of a {\em I}~=~23.7 mag galaxy.

{\bf J123612+621138/J123612+621140} (Starburst Candidates): are
separated by $\sim4$~arcsec.  J123612+621140 has a steep radio
spectrum ($\alpha>$0.89). J123612+621138 has no spectral index
measurement although the radio sources are only partially separated by
Richards (2000) who classifies the pair as a single object.  Each one
appears to be an extended starburst and is identified with a separate
galaxy in a system which appears to be interacting. J123612+621138 is
identified with an {\em I}~=~18.9 mag galaxy; J123612+621140 is
identified with an {\em I}~=~19.4 mag galaxy. Cowie et al.\/ (2004)
measure the same redshift for each galaxy ($z=0.275$). J123612+621138
is detected by {\em Chandra}.

{\bf J123615+620946} (Starburst Candidate): is extended over
$\sim$0.8~arcsec and overlies the nuclear region of an {\em I}~=~22.0
mag galaxy with a redshift of 1.263 (Cowie et al.\/ 2004).

{\bf J123616+621513} (Starburst Candidate): is extended over
$\sim$1~arcsec. The weak extended emission $\sim$3~arcsec to the NW is
only visible at low resolution and has no optical counterpart. It is
not thought to be associated with J123616+621513 and falls below the
7$\sigma$ detection threshold for the 10-arcmin field sample.
J123616+621513 is identified with an {\em I}~=~22.9 mag
galaxy. Chapman et al.\/ (2004b) identify the system as a sub-mm galaxy
(S$_{850}=5.8\pm1.1$ mJy) with a redshift of 3.61.  It is detected by
{\em Chandra}.

{\bf J123617+621011} (Unclassified): is extended over $\sim$0.6~arcsec
and contains a compact component at 0.2-arcsec resolution (not
shown). The nature of the object is unclear since the radio structure
could originate from AGN or starburst activity (or both). It is
identified with a {\em I}~=~21.6 mag galaxy. Hornschemeier et al.\/
(2001) find a single emission line at 6879.3 {\AA} which they assign
as [OII] at a redshift of 0.845.  It is detected by {\em Chandra}
with a X-ray luminosity strongly suggestive of AGN activity.

{\bf J123617+621540} (Unclassified): has a  fairly steep radio
spectrum ($\alpha>$0.55). It contains a compact component with
two-sided emission extended over 1.3~arcsec. It is identified with a
faint galaxy ({\em I}~=~23.9 mag) in a complex region on the CFHT
frame.

{\bf J123618+621635} (AGN Candidate): contains two extended radio components
separated by 2.5~arcsec. The radio component to the SE overlies the
central region of an {\em I}~=~19.7 mag galaxy and shows no evidence
for any compact core.  The NW radio component appears to be associated
with a second galaxy which may be part of an interacting
system. Hornschemeier et al.\/ (2001) find two emission lines from the
SE component which they assign to be [OIII] at a redshift of 0.679.
It is also detected by {\em Chandra}.  On the basis of the X-ray
luminosity and a weak [NeV] feature, Hornschemeier et al.\/ (2001)
suggest that this object is an AGN system.

{\bf J123618+621550} (Unclassified): exhibits a steep radio spectrum
($\alpha>$0.63) compact component with one-sided emission extended
over 0.4 arcsec.  It is unidentified at the limit of the CFHT
frame. Chapman et al.\/ (2004b) identify the system as a possible
sub-mm galaxy (S$_{850}=7.3\pm1.1$ mJy) with a redshift of
1.87. However, this identification is uncertain since the measured
redshift is for the optical object approximately 0.5 arcsec to the SW
of the radio source; this is above the nominal astrometric alignment
errors for the position of the radio source in the 10-arcmin field.

{\bf J123619+621252} (Starburst Candidate): has a steep radio spectrum
($\alpha>$0.8) and has  radio emission extended over 0.4~arcsec. It
lies at the nucleus of an {\em I}~=~19.9 mag galaxy at
redshift~=~0.473 (Cowie et al.\/ 2004). The optical galaxy appears to
have a companion and may be part of a merging system.  It is detected
by {\em Chandra}.

{\bf J123620+620844} (AGN): has a flat radio spectrum ($\alpha$~=~0.01) and
shows a compact core with a one-sided (0.6~arcsec) extension to the
east. The core component overlies the nucleus of an {\em I}~=~21.5 mag
galaxy at a redshift of 1.0178 (Cowie et al.\/ 2004).

{\bf J123621+621109} (Starburst Candidate): has a steep radio spectrum
($\alpha>$0.86).  The VLA-only image (not shown) reveals additional low surface
brightness emission extending over 3~arcsec to the north-east. A
brighter but still extended component is seen in the combination image
and overlies the fainter, south-western member of an {\em I}~=~21.8
mag pair of merging galaxies at a redshift of 1.014 (Cohen et al.\/
2000).  There is a {\em Chandra} counterpart.

{\bf J123621+621708} (Unclassified): has a steep radio spectrum
($\alpha>$0.56) and contains a compact component with one-sided
emission extending to the south-east. An extended radio component lies
2.5~arcsec to the north-west, close to the position of an {\em
I}~=~23.4 mag galaxy on the CFHT frame.  There is no optical object
brighter than {\em I}~=~25 mag at the position of the compact radio
component. Chapman et al.\/ (2004b) identify the system as a possible
sub-mm galaxy (S$_{850}=7.8\pm1.9$ mJy) with a redshift of
1.99. However, this identification is uncertain since the measured
redshift is for the galaxy to the NW of the radio source; this is well
above the nominal astrometric alignment errors for the position of the
radio source in the 10-arcmin field.

{\bf J123622+621544} (AGN Candidate): has a steep radio spectrum
($\alpha>$0.6). The radio structure is extended and low
surface-brightness emission seen in the VLA-only image (not shown)
extends for nearly 3~arcsec to the north-west.  A brighter region of
emission region lies 0.75~arcsec to the south-east of the bright
nucleus of an {\em I}~=~19.8 mag spiral galaxy in the HFF at a
redshift of 0.6471 (Cowie et al.\/ 2004). As the radio emission extends
well beyond the optical galaxy the source is classified as an
AGN candidate with an undetected core. There is a faint {\em Chandra}
counterpart.

{\bf J123622+621629} (Starburst Candidate): is extended (1.3~arcsec)
and is identified with an {\em I}~=~23.8 mag galaxy on the CFHT
frame. Chapman et al.\/ (2004b) identify the system as a sub-mm galaxy
(S$_{850}=7.7\pm1.3$ mJy) with a redshift of 2.40. There is a faint {\em
Chandra} counterpart

{\bf J123622+620945} (Starburst Candidate): is extended over
$\sim$0.8~arcsec and shows evidence for double structure at position
angle 110 degrees. There is no evidence of any compact radio structure
at full resolution. The source overlies the nuclear region of an {\em
I}~=~21.1 mag galaxy with a redshift of 0.7479 (Cowie et al.\/ 2004).

{\bf J123623+621642} (AGN Candidate): has a steep radio spectrum
($\alpha$~=~0.63) and contains a compact component together with
one-sided extended emission. It is identified with an {\em
I}~=~23.9 mag galaxy in a complex region on the CFHT frame.

{\bf J123624+621017} (Unclassified): consists of two extended
components separated by 2.1~arcsec at position angle 30\degr. Each
component is $\sim$0.5~arcsec in size and there is no evidence for any
compact radio structure at 0.2-arcsec resolution. The source cannot be
identified with any optical object brighter than {\em I}~=~25 mag in the HFF
frame.

{\bf J123624+621743} (Unclassified): shows a compact feature in the
0.2 arcsec image (not shown) with additional emission extending over
0.9~arcsec. It is identified with a {\em I}~=~22.9 mag galaxy and has
a faint {\em Chandra} counterpart.

{\bf J123629+621045} (Unclassified): has a steep radio spectrum with
$\alpha>0.86$ and is extended east-west over 2.7~arcsec with no
evidence of compact emission at the highest angular resolution.  The
eastern end of the radio source is coincident with the nucleus of an
{\em I}~=~22.2 mag galaxy at a redshift of 1.013 on the CFHT frame
(Cohen et al.\/ 2000) but the extension is not correlated with optical
emission.  Chapman et al.\/ (2004b) identify this system as a sub-mm
source (S$_{850}=5.0\pm1.3$ mJy). It is detected by {\em Chandra}.

{\bf J123630+620923} (Starburst Candidate): is extended over
0.6~arcsec and overlies the nuclear region of an {\em I}~=~22.5 mag
galaxy on the CFHT frame with an estimated photometric redshift of
0.953 (Bauer et al.\/ 2002).  It is detected by {\em Chandra}.

{\bf J123630+620851} (Starburst Candidate): is extended over
3.3~arcsec. The extension is more clearly seen in the VLA-only image
(not shown) and additional flux is seen with the Westerbork radio
telescope indicating possible emission on the 10-arcsec scale.
The radio structure roughly follows the optical isophotes of a complex
galaxy system that appears to be interacting, the brightest member of
which has {\em I}~=~22.6 mag.

{\bf J123631+620957} (Unclassified): has a steep radio spectrum
($\alpha>0.99$) and a compact component with two-sided emission
extending over 0.8~arcsec. There is no optical counterpart brighter
than {\em I}~=~25 mag on the HFF frame but it has a faint {\em
Chandra} counterpart.

{\bf J123632+621658} (Starburst Candidate): is extended over
$\sim$4~arcsec on the VLA-only image (not shown) and shows no sign of
compact emission at higher resolution. The source is very heavily
resolved with MERLIN and the position in Table~\ref{10arcmin.tab} is
taken from the VLA-only image at 2-arcsec resolution.  The source is
associated with an {\em I}~=~18.2 mag spiral galaxy with a redshift of
0.437 (Cowie et al.\/ 2004).

{\bf J123632+620759} (Unclassified): has a steep radio spectrum
($\alpha>0.58$) and is extended over some 2.6~arcsec on the VLA-only
image (not shown). At higher resolution a
compact component is found with two-sided extended emission, although
the majority of the extended emission seen in the VLA-only image is
resolved out. The radio source lies outside the CFHT frame so no
optical image is shown in Fig.~\ref{colourpix}. However Cowie et al.\/
(2004) catalogue an {\em R} = 24.0 mag galaxy at the position of the radio
source with a redshift of 1.9939. It is detected by {\em Chandra}.

{\bf J123633+621005} (Starburst Candidate): has a steep radio spectrum
($\alpha>0.98$) and is extended over approximately 1.2~arcsec.  There
is no evidence for any compact emission at the highest angular
resolution.  It overlies a face-on {\em I}~=~21.6 mag spiral galaxy at
a redshift of 1.016 (Cowie et al.\/ 2004) on the HFF frame. It is
detected by {\em Chandra}.

{\bf J123634+621213} (Starburst): has a steep radio spectrum
($\alpha$~=~0.74). The extended radio emission is two-sided,
surrounding a compact component.  However EVN observations at 20 mas
resolution do not detect this compact component (Garrett et al.\/
2001). The source is identified with a bright disc galaxy ({\em
I}~=~18.7 mag), at a redshift of 0.456 (Cowie et al.\/ 2004).  The HFF
optical image shows evidence for a dust-lane or possible double
nucleus which may indicate a recent merger. The brightest radio
emission lies between the optical nuclei and the extensions follow the
optical isophotal ridge-line.  The galaxy is also detected by {\em
ISO} with $S_{15\mu \rm m} = 448~\mu$Jy. X-ray emission is detected by
{\em Chandra}.

{\bf J123634+621241} (Starburst): has a steep radio spectrum 
($\alpha$ ~=~ 0.74).  There is extended emission to the south-west of a
compact component which is not detected by the EVN at 20 mas
resolution (Garrett et al.\/ 2001). The compact component overlies the
nuclear region of an {\em I}~=~22.3 mag Scd galaxy at a redshift of
1.219 on the HFF WFPC2 frame (Cohen et al.\/ 2000). It has an mid-IR
flux density $S_{15\mu \rm m} = 363~\mu$Jy as detected by {\em ISO}, which
implies a substantial star-formation rate.  This source is the most
likely radio counterpart to SCUBA HDF850.7 (see
Section~\ref{sect-HDF850.1}). There is a faint {\em Chandra}
counterpart.

{\bf J123635+621424} (Starburst + AGN?): has a steep radio spectrum
($\alpha>0.87$).  It appears relatively compact, with an angular size
of only $\leq$0.3~arcsec, in contrast to the earlier angular size
estimate of 2.8~arcsec reported by Richards et al.\/ (1998).  The
radio source overlies the bright optical nucleus of a face-on {\em
I}~=~22.6 mag spiral galaxy on the HFF frame. The spatial extent of
the galaxy suggests that it is of low to moderate redshift.  However,
Bauer et al.\/ (2002) list this object as having a spectroscopic
redshift close to 2 and Dawson et al.\/ (2002) assign a redshift of
2.011 from several detected emission lines. It is an {\em ISO}
detection and Chapman et al.\/ (2004b) identify the system as a sub-mm
galaxy (S$_{850}=5.5\pm1.4$ mJy) at a a redshift of 2.01. It is
detected by {\em Chandra}.  Dawson et al.\/ (2002) argue that the
15~$\mu$m and X-ray detections are indicative of an AGN system.  If
this object is a compact nuclear starburst system, the implied
star-formation rate from the radio and IR flux densities is very large
($\sim$1800 M$_{\odot}$ yr$^{-1}$).

{\bf J123636+621320} (Starburst Candidate): is extended over
approximately 0.7~arcsec and shows no sign of compact emission. There
is no optical counterpart brighter than {\em I}~=~25 mag at the radio
source position on the HFF frame.  There is a faint {\em Chandra}
counterpart.

{\bf J123637+620852} (Starburst Candidate): has a steep radio spectrum
($\alpha>0.82$), is extended east-west over approximately 1.1~arcsec 
and shows no sign of a compact component. It is identified with an
{\em I}~=~24.5 mag galaxy in a complex region on the CFHT frame.

{\bf J123640+621009} (AGN Candidate): consists of two components
separated by 1.2~arcsec. The source as a whole has a radio
spectrum of $\alpha$ = 0.44, which suggests the presence of a
flat-spectrum component. The 8.4-GHz position (Richards et al.\/ 1998)
is close to the northerly component indicating that it has a flat
spectrum and therefore may be associated with an AGN. Although no
evidence for a compact feature can be found at 1.4 GHz, the radio
image has a low signal-to-noise ratio.  The more northerly component
overlies a faint compact {\em I}~=~23.9 mag galaxy on the HFF frame.

{\bf J123641+620948} (Unclassified): has a steep radio spectrum
($\alpha>0.56$) and contains a compact radio component with some
extended emission with a total angular extent of approximately
0.6~arcsec.  The compact component coincides with the nucleus of an
{\em I}~=~20.2 mag disc galaxy at a redshift of 0.518 which may be
part of a merging system (Cohen et al.\/ 2000). The extended radio
emission lies perpendicular to the major axis of the optical galaxy.
There is a faint {\em Chandra} counterpart.

{\bf J123642+621331} (Starburst + AGN?): has a steep radio spectrum
($\alpha$ = 0.94).  The radio structure consists of a compact component
together with emission extended over 0.4~arcsec. Garrett et al.\/
(2001) detect the dominant compact component with the EVN at an
angular resolution of 20 mas; it may therefore be a steep-spectrum
AGN core.  There is no optical counterpart at the detection threshold in
the HFF {\em I}-band image. Waddington et al.\/ (1999) detect a very
red object at the radio source position with NICMOS, in J and H-band,
and with the KPNO 4-m in {\em K}-band. On the basis of a single Ly$\alpha$
emission line and the observed spectral energy distribution, they
ascribe a redshift of 4.424 and argue that it is a distant starburst
galaxy with a weak embedded AGN. This interpretation is consistent
with the radio morphology. The source is detected by both {\em ISO}
and {\em Chandra}.

{\bf J123642+621545} (Unclassified): has an intermediate radio
spectrum ($\alpha=0.50$). The radio structure consists of a
compact component and extended emission to the NE and SW with the
latter reaching nearly 1.5~arcseconds from the core.  The core
component overlies the nucleus of an {\em I}~=~20.7 mag galaxy on the
CFHT frame at a redshift of 0.857 (Hornschemeier et al.\/ 2001).  The
source is detected by both {\em ISO} and {\em Chandra}; the compact
component could be an AGN, but the nature of this object remains
unclear.

{\bf J123644+621133} (AGN -- FRI radio galaxy): has a bright core with
a flat radio spectrum ($\alpha$ = 0.1, Richards et al.\/ 1998) with
steep spectrum emission oriented N--S and extending for about
15~arcsec in the VLA-only image (not shown). Overall the radio source
exhibits a classical FRI structure.  It is optically identified with
an {\em I}~=~20.9 mag elliptical galaxy at a redshift of 1.050 (Cohen
et al.\/ 2000). This source is not catalogued by Aussel et al. (1999)
but Goldschmidt et al. (1997) list a nearby {\em ISO} detection at 6.7
$\mu$m only; Mann et al. (1997) identify this with the host of
J123644+621133.  There is a faint {\em Chandra} counterpart.

{\bf J123645+620754} (Unclassified): lies close to the edge of the
10-arcmin field. The source is extended east-west over some 3~arcsec,
and there is no evidence for any compact emission. The position given
in Table~\ref{10arcmin.tab} is from the VLA-only image at 2-arcsec
resolution.  The radio source lies outside the CFHT frame so no
optical image is shown in Fig.~\ref{colourpix}. However Cowie et al.\/
(2004) catalogue an {\em R} = 23.0 mag galaxy at the position of the
radio source with a redshift of 1.433. There is a faint {\em Chandra}
counterpart.

{\bf J123646+621448}~(Unclassified): has a steep radio spectrum,
($\alpha>0.84$) and contains a compact component with a one-sided
0.5~arcsec extension at a position angle of $\approx$160 degrees. It is not
identified on the HFF frame to approximately {\em I}~=~25
mag. However, on the lower resolution CFHT frame it overlies a faint
but distinct galaxy ({\em I}~=~25.5 mag).  Chapman et al.\/ (2004b)
identify this system as a sub-mm source (S$_{850}=10.3\pm2.2$ mJy).
There is a faint {\em Chandra} counterpart.

{\bf J123646+621629} (Starburst Candidate): has a very steep radio spectrum
($\alpha>1.62$).  The radio structure consists of a compact central
feature together with extended emission distributed symmetrically
around it with a total angular extent of about 2.4~arcsec.  The radio
core is coincident with the nuclear region of an {\em I}~=~19.8 mag
galaxy at a redshift of 0.502 (Cohen et al.\/ 2000).

{\bf J123646+621404} (AGN): has an inverted radio spectrum 
($\alpha$=-0.04).  The radio structure has a compact component
 and two-sided extended emission.  The radio emission overlies the
nucleus of a nearly face-on disc galaxy ({\em I}~=~20.7 mag) at a
redshift of 0.961 (Cowie et al.\/ 2004), which exhibits a broad
emission line spectrum (Phillips et al.\/ 1997). These observations
combine to form strong evidence that this galaxy contains an AGN. This
radio source was detected by {\em ISO} with $S_{7\mu \rm m}$ ~=~ 52
$\mu$Jy and it has a bright {\em Chandra} counterpart. Although
Rowan-Robinson et al.\/ (1997) interpret this source as a massive
starburst, star-formation seems unlikely to be powering the bulk of
the radio emission in this object.
	
{\bf J123646+620833} (Starburst Candidate): has a steep radio spectrum
($\alpha>0.61$). The radio structure is extended over 0.8~arcsec and
shows no sign of compact emission at 0.2-arcsec resolution. The radio
emission overlies the nuclear region of an {\em I}~=~20.6 mag galaxy
at a redshift of 0.9712 (Cowie et al.\/ 2004).  There is a faint {\em
Chandra} counterpart.

{\bf J123646+621226} (Unclassified): has a steep radio spectrum
($\alpha>1.14$) and is completely resolved-out at 0.2-arcsec
resolution. The VLA-only image (as shown in Fig.~\ref{scubapix}) shows
the source to be 3.2~arcsec in extent. No optical
counterpart brighter than {\em I}~=~28 mag is found at the radio
position. Deep near-infrared observations (Thompson et al.\/ 1999)
place magnitude limits of $I_{\rm AB}>$28.5, $J_{\rm 110}>$29, and
$H_{\rm 160}>$29 mag.

{\bf J123646+621445} (Unclassified): has a steep radio spectrum
($\alpha = 0.98$) and is extended over 2.3~arcsec with the major axis
roughly aligned with J123646+621448 which lies 6~arcsec to the north
west. However, J123646+621445 overlies a fairly compact faint
(I$\sim$23.9 mag) galaxy. Since both sources overlie separate optical
identifications, it is most likely that they are unrelated. The {\em
  I}~=~20.4 mag galaxy 2.5~arcsec to the NE of J123646+621445 has a
measured redshift of 0.558 (Cohen et al.\/ 2000), but does not seem to
be related to the radio source. The {\em ISO} source at 4 arcsec
separation in a similar direction cannot therefore be identified with
the radio source with any confidence.

{\bf J123649+621313} (Starburst): has a steep radio spectrum 
($\alpha$ = 0.72) with no evidence of any compact structure at
0.2-arcsec resolution.  It is extended by approximately 1~arcsec along
the major axis of an {\em I}=21.1 mag disc galaxy at a redshift of
0.475 (Cohen et al.\/ 2000). The galaxy is in the {\em ISO} catalog of
HDF sources with $S_{15 \mu \rm m}$ = 320 $\mu$Jy and may be associated
with SCUBA HDF850.4 (see Section~\ref{sect-HDF850.4}). There is a faint
{\em Chandra} counterpart.  Low level radio emission below the formal
detection threshold (and hence not in the catalogue) may also be seen
from an {\em I}~=~22.7 mag disc galaxy located about 3~arcsec to the
north which lies at a redshift of 1.238 (Cohen et al.\/ 2000).
	
{\bf J123650+620801} (Starburst Candidate): is extended by approximately
0.7~arcsec and shows no sign of compact structure at 0.2-arcsec
resolution. The radio source lies outside the CFHT image
so no optical frame is shown in Fig.~\ref{colourpix}. However Cowie et
al.\/ (2004) catalogue an {\em R} = 22.2 mag galaxy at the position of the
radio source with a redshift of 0.559.

{\bf J123650+620844} (Starburst Candidate): has a steep spectrum
($\alpha>0.80$).  The radio structure shows a compact component
together with extended emission of approximately 0.8~arcsec extent.
The compact component overlies the nucleus of an {\em I}~=~20.0 mag
galaxy at a redshift of 0.434 (Cowie et al.\/ 2004).

{\bf J123651+621030} (Starburst): has a steep-spectrum
($\alpha=0.74$). The radio source is extended over 1.2~arcsec and
overlies the nuclear region of an {\em I}~=~19.9 mag disc galaxy at a
redshift of 0.410 (Cohen et al.\/ 2000).  It is detected by {\em ISO}
with $S_{15 \mu \rm m}$ = 341 $\mu$Jy.  There is a faint {\em Chandra}
counterpart.

{\bf J123651+621221} (Starburst + AGN?): has a steep spectrum
($\alpha>0.71$). The MERLIN+VLA image shows a source extended over
$\sim$1.2~arcsec with no evidence for any compact component. Garrett
et al.\/ (2001) also find no evidence for any compact component with
the EVN at an angular resolution of 20 mas. The radio structure thus
favours a starburst rather than an AGN interpretation.
The radio source is identified with very faint emission (I$>$28 mag)
on the HDF frame. Dickinson et al.\/ (2000) and Dickinson (2000)
report a very red object ($I_{\rm AB}$~=~27.8 mag, $J_{\rm 110}-
H_{\rm 160}$~=~1.6) at the position of this source from NICMOS
observations, implying a dust-obscured starburst system at a redshift
approaching 3.  The source is detected by {\em ISO} with $S_{15 \mu
\rm m}$ = 48 $\mu$Jy.  Hornschemeier et al.\/ (2001) find a hard X-ray
detection with {\em Chandra} suggesting an obscured QSO at a redshift
of 2.7. 

The bright ({\em I}~=~21 mag) galaxy lying 1.5~arcsec to the south is
at a redshift of 0.401 (Cohen et al.\/ 2000), and is not thought to be
related to the radio source.

{\bf J123652+621444} (AGN): has an inverted spectrum ($\alpha = -0.12$)
(Richards et al.\/ 1998) and is observed to vary in intensity on the
time scale of months, suggesting the presence of an AGN (Richards
2000). The radio structure shows a compact core and one sided jet-like
feature to the east $\sim1.3$~arcsec in length.  The low resolution WSRT
image of this source (Garrett et al.\/ 2000) shows evidence for low
surface-brightness emission extending over $\sim$30~arcsec to the
east. If this is an extension to the core-jet structure seen in the
high resolution MERLIN+VLA image, the jet has an overall extent of
$\sim$100 kpc. The core overlies  the nucleus of an {\em I}~=~18.5 mag
elliptical galaxy on the HDF frame at a redshift of 0.321 (Cowie et
al.\/ 2004). It is detected by {\em Chandra}.

{\bf J123653+621139} (Starburst): has a steep spectrum
($\alpha>0.77$).  The radio structure shows a compact component with a
slight ($\sim0.3$-arcsec) extension to the east.  The compact
component overlies the nucleus of an {\em I}~=~21.9 mag compact galaxy
at a redshift of 1.275 (Cohen et al.\/ 2000). It is also detected by
{\em ISO} at 15 $\mu$m with a flux density of 180 $\mu$Jy.  There is a
faint {\em Chandra} counterpart.

{\bf J123654+621040} (Starburst Candidate): has a steep spectrum
($\alpha>1.0$) and is extended over approximately 1.5~arcsec; it shows no
signs of compact emission at 0.2-arcsec  resolution. The source lies
close to the edge of the HFF frame but there is no optical
identification at the radio source position brighter than {\em I}~=~25 mag.

{\bf J123655+620917} (Unclassified): has a radio spectrum with
$\alpha>0.55$.  The source contains a compact component associated
with the galaxy nucleus and emission extending over 0.6~arcsec. The
compact component overlies the nucleus of an {\em I}~=~20.4 mag galaxy
at a redshift of 0.419 (Cowie et al.\/ 2004)

{\bf J123655+620808} (Unclassified): has a steep radio spectrum
($\alpha>0.85$) and shows a compact component with emission extending
over 0.7~arcsec. The radio source lies outside the CFHT image so no
optical frame is shown in Fig.~\ref{colourpix}. However Cowie et al.\/
(2004) catalogue an {\em R} = 22.6 mag galaxy at the position of the
radio source with a redshift of 0.792. There is a faint {\em Chandra}
counterpart.

{\bf J123656+621207} (Starburst Candidate): has a steep spectrum
($\alpha>1.32$), is extended over $\sim1$~arcsec and shows no sign of
any compact emission at 0.2-arcsec resolution. It lies close to the
edge of an HDF frame and at the radio position there is no optical
object brighter than {\em I}~=~28 mag. Richards et al.\/ (1999) report
that NICMOS imaging shows no counterpart to the limits {\em J} $>25$
and {\em H} $>25$; Barger et al.\/ (1999) find no counterpart at {\em
K} $>24$. The radio source is located $\sim$3~arcsec from the second
brightest SCUBA source, HDF850.2 (Hughes et al.\/ 1998), but the
association is uncertain -- see the detailed discussion in
Section~\ref{sect-HDF850.2}.

{\bf J123656+621301} (Starburst): has a steep radio spectrum
($\alpha>1.22$), is extended over 1.2~arcsec and is identified with an
{\em I}~=~22.5 mag compact galaxy at a redshift of 0.474 (Bauer et
al.\/ 2002). It has an {\em ISO} counterpart and is detected by {\em
Chandra}. The bright companion galaxy to the SE has a redshift of
0.475 (Cohen et al.\/ 2000).

{\bf J123659+621449} (Starburst Candidate): is extended over 1.0~arcsec and
shows no sign of compact emission at 0.2-arcsec resolution. The radio
structure overlies a disturbed {\em I}~=~20.9 mag disc galaxy at a
redshift of 0.761 (Cowie et al.\/ 2004). The galaxy appears to show a
dust-lane or double nucleus and may be undergoing a merger.  The
source is detected by {\em ISO} and has a faint {\em Chandra} counterpart.

{\bf J123700+620909} (AGN Candidate): has a steep radio spectrum
($\alpha$ = 0.89).  There is a compact component with one-sided
emission to the NE; the total angular size of the source being
$\sim$0.4~arcsec. The radio emission There is no optical counterpart
at the radio position to the limit of {\em I}~=~25 mag in the CFHT
frame.

{\bf J123701+621146} (Starburst): has a steep radio spectrum
($\alpha$~=~0.67) and is extended over 3~arcsec in the VLA-only image
at 2-arcsec resolution (not shown). At 0.5-arcsec resolution some of
the outer low surface brightness emission is resolved out, but the
central 1.5~arcsec of emission is well imaged and shows no sign of any
compact component. There is no optical detection at the radio position
on the HFF frame brighter than {\em I}~=~25 mag.  The source is
detected by {\em ISO}.  J123701+621146 is the most likely radio
counterpart to SCUBA 850.6 (see Section~\ref{sect-HDF850.6}). There is a
faint {\em Chandra} counterpart.

{\bf J123702+621401} (Unclassified): has a steep radio spectrum
($\alpha> 0.79$) and is extended over 3.4~arcsec in the VLA-only image
(not shown). At 0.5-arcsec resolution the source shows no sign of
any compact component and some of the low surface brightness emission
is resolved out. The SE end of the source overlies an {\em I}~=~21.9
mag galaxy at a redshift of 1.2463 (Cowie et al.\/ 2004), with a
significant fraction of the extended emission lying well outside the
confines of the optical galaxy; the latter precludes a straightforward
classification as a starburst candidate. It is detected by {\em ISO}.

{\bf J123704+620755} (Starburst Candidate): is extended over 1~arcsec
and shows no sign of any compact component. It is outside the  CFHT frame but it is detected by {\em
Chandra}.

{\bf J123705+621153} (Starburst): has a steep spectrum ($\alpha>1.27$)
and is extended over 1.5~arcsec with no evidence for any compact
component at 0.2-arcsec resolution. The source is centred on the
nuclear region of an {\em I}~=~20.6 mag face-on spiral galaxy at a
redshift of 0.902 (Cowie et al.\/ 2004).  The source is detected by
{\em ISO} and has a faint {\em Chandra} counterpart.

{\bf J123707+621408} (Starburst Candidate): has a relatively flat
radio spectrum ($\alpha=0.29$). The source is extended over 0.9~arcsec
and there is no evidence for any compact component.  The radio
morphology favours a starburst classification and the relatively flat
spectrum may be attributed to low-frequency free-free
absorption. There is no obvious optical identification on the HFF
frame although there does appear to be a faint {\em I}~=~24.9 mag
galaxy at the radio position on the CFHT frame. Chapman et al.\/
(2004b) identify the system as a sub-mm galaxy (S$_{850}=4.7\pm1.5$
mJy) with a redshift of 2.48.  It is detected by {\em Chandra}.

{\bf J123707+621121} (Starburst Candidate): has a steep radio spectrum
($\alpha>0.82$). It is extended over 0.6~arcsec and there is no
evidence for any compact component. There is a faint {\em I}~=~25 mag
object at the radio source position on the CFHT frame.

{\bf J123708+621056} (Starburst Candidate): has a relatively flat
radio spectrum ($\alpha$ = 0.35) but there is no evidence for any
compact radio component at 0.2-arcsec resolution. The radio spectrum
may therefore be due to substantial free-free absorption at 1.4 GHz.
The source is extended over 0.7~arcsec and is displaced by 
$\sim$0.5~arcsec from the central region of an {\em I}~=~19.8 mag
edge-on disc galaxy at a redshift of 0.422 (Cowie et al.\/ 2004). This
separation is probably real since the registration errors, even
towards the edge of the HFF, are substantially lower than 0.5~arcsec.
There is a faint {\em Chandra} counterpart.

{\bf J123709+620837/J123709+620841} (AGN and AGN Candidate): are
separated by 4.5~arcsec. J123709+620837 has a flat radio spectrum
($\alpha = -0.35$). J123709+620841 is undetected at 8.4 GHz implying a
steeper radio spectrum, although it is difficult to separate these
sources in the 8.4-GHz image where the angular resolution is
comparable with the source separation. They both have compact cores
and faint extended emission. The compact component dominates the radio
emission in J123709+620837, which is unambiguously identified as an
AGN system with a compact flat-spectrum core. J123709+620841 may also
be an AGN system, though this is less clear.  Each radio source is
identified with a separate galaxy ({\em I}~=~20.4 and 21.4 mag with
redshifts of 0.907 and 0.902 respectively (Cowie et al.\/ 2004); in
each case the compact radio component  overlies the nucleus of
the galaxy.  Both sources have {\em Chandra} counterparts,
J123709+620837 has the brighter X-ray emission.

{\bf J123711+621330} (AGN Candidate): has a steep radio spectrum
($\alpha$ = 0.69). The source contains a compact radio component with
one-sided jet-like emission extending $\sim1.2$ arcsec to the east.
The compact component is coincident with the brightest part of an {\em
I}~=~22.7 mag galaxy.  The optical galaxy is asymmetric in the same
sense as the radio structure. Bauer et al.\/ (2002) estimate a
photometric redshift of 1.112 for this galaxy. The classification as
an AGN candidate is based on the registration of the compact component
with the bright optical nucleus together with the asymmetry of the
radio structure. The associated X-ray emission detected by {\em
Chandra} is significantly offset by 0.9 arcsec to the east of
the radio peak.

{\bf J123711+621325} (Starburst Candidate): has a steep radio spectrum
($\alpha>1.16$) and extends over 0.8~arcsec and shows no sign of
compact emission at 0.2-arcsec resolution.  The source is identified
with an {\em I}~=~25 mag galaxy on the CFHT frame. Chapman et al.\/
(2004b) identify the system as a sub-mm galaxy (S$_{850}=4.2\pm1.3$
mJy) with a redshift of 1.99. There is a faint {\em Chandra}
counterpart.

{\bf J123713+621603} (Starburst Candidate): is extended over
0.9~arcsec with no evidence for any compact radio component at
0.2-arcsec resolution. It is identified with an {\em I}~=~22.8 mag
galaxy on the CFHT frame; the source overlies the nuclear region.

{\bf J123714+621558} (Starburst Candidate): is extended over 1.5~arcsec
with  no sign of any compact component at 0.2-arcsec resolution. The
source is identified with an {\em I}~=~20.8 mag galaxy on the CFHT frame at a
redshift of 0.567 (Cohen et al.\/ 2000).

{\bf J123714+620823} (AGN): has an inverted radio spectrum
($\alpha = -0.15$) and the radio structure is almost unresolved. The
image shown in Fig.~\ref{colourpix} and described in
Tables~\ref{10arcmin.tab} and~\ref{10arcminchars} is from the MERLIN
data only. It is identified with the nucleus of an {\em I}~=~22.0 mag
galaxy on the CFHT frame.

{\bf J123716+621512} (AGN Candidate): has a relatively flat radio
 spectrum ($\alpha$~=~0.41).  The source consists of a compact
 component and one-sided emission extending $\sim0.4$ arcsec to the
 south-west.  The source overlies the nucleus of an {\em I}~=~19.8 mag
 galaxy on the CFHT frame with a redshift of 0.230 (Cowie et al.\/
 2004).  There is a faint {\em Chandra} counterpart.

{\bf J123716+621643} (Starburst Candidate): is extended over
0.9~arcsec.  The radio emission overlies what appears to be a pair of
merging galaxies on the CFHT frame.  The radio emission is offset from
the brighter galaxy by $\sim$1~arcsec. This is real since the
registration errors in this region are $<$250 mas.  The brighter
galaxy has {\em I}~=~21.6 mag and a redshift of 0.557 (Cohen et al.\/
2000).There is a faint {\em Chandra} counterpart.

{\bf J123716+621733} (AGN Candidate): has a steep radio spectrum
($\alpha>0.76$). The radio structure contains a compact component with
a $\sim$0.6~arcsec one-sided extension to the S-W. The compact
component overlies the nucleus of an {\em I}~=~22.2 mag galaxy with a
redshift of 1.146 (Cowie et al.\/ 2004). It is detected by
{\em Chandra}.

{\bf J123716+621007} (Starburst Candidate): has emission extending over
$\sim$0.7~arcsec. It overlies the nucleus of an {\em I}~=~19.9 mag
galaxy at a redshift of 0.411 (Cohen et al.\/ 2000).  There is a faint
{\em Chandra} counterpart.

{\bf J123717+620827} (Starburst Candidate): has a steep radio spectrum
 ($\alpha>0.77$). The radio structure is extended N-S over
2.1~arcsec.  At 0.5-arcsec resolution some of the low surface
brightness emission is resolved out and there is no evidence for any
compact component. The brightest part of the radio structure overlies
the nuclear region of an {\em I}~=~19.7 mag galaxy on the CFHT frame.

{\bf J123719+620902} (Starburst Candidate): is extended over
$\sim$0.7~arcsec and overlies the nuclear region of an {\em I}~=~22.1
mag galaxy on the CFHT frame. This source is not listed in Richards
(2000).

{\bf J123721+621129} (AGN): has an inverted spectrum ($\alpha$ =
--0.28) with a dominant compact radio component and one-sided emission
extending $\sim$0.8~arcsec to the north. This source is thus almost
certainly an AGN system. There is no optical identification at the
radio source position brighter than {\em I}~=~25 mag. There is a
distorted disc galaxy to the NE with an estimated photometric redshift
of 2.468 (Bauer et al.\/ 2002), but the radio-optical offset of over
1~arcsec is far too high to be accounted for by astrometric
errors. Richards et al.\/ (1999) report that from deep {\em I} and {\em
  K} band imaging (Barger et al.\/ 1999), this radio source is
associated with a very red object with ($I-K>$5.2).  There is a faint
{\em Chandra} counterpart.

{\bf J123721+621346} (Starburst Candidate): has a steep radio spectrum
($\alpha>0.66$). The radio structure is extended over 0.7~arcsec and
shows no evidence for any compact component at 0.2-arcsec resolution.
The source is identified with an {\em I}~=~22.1 mag galaxy at a
redshift of 1.019 (Cohen et al.\/ 2000).

{\bf J123725+620856} (Unclassified):  is extended over
$\sim$0.4~arcsec. It overlies the nuclear region of an
asymmetrical {\em I}~=~22.1 mag galaxy, with an estimated photometric
redshift of 0.984 (Bauer et al.\/ 2002).  It is detected by {\em Chandra}.

{\bf J123725+621005} (Unclassified): is extended over $\sim$3.8~arcsec
aligned roughly E-W. There is no evidence for any compact component
and at 0.2--0.5-arcsec angular resolution some of the low surface
brightness emission has been resolved out. The position given in
Table~\ref{10arcmin.tab} is from the VLA-only image (not shown). The
radio source overlies the nuclear region of an {\em I}~=~19.1 mag
galaxy with an estimated photometric redshift of 0.360 (Bauer et al.\/
2002).

{\bf J123725+621128} (AGN~--~WAT~radio~galaxy): has a steep radio
spectrum ($\alpha$~=~1.35). The radio lobes extend over
$\sim$3.8~arcsec and are oriented E-W.  These lobes are characteristic
of a classical wide-angled tail (WAT) radio structure.  The compact
radio core overlies the nucleus of a compact, {\em I}~=~22.9 mag
galaxy and

{\bf J123730+621258} (Unclassified): has a steep spectrum
($\alpha>0.79$) and contains a compact radio component with extended
emission over $\sim$0.3~arcsec. The object lies outside the area
of the CFHT frame.

{\bf J123732+621012} (Unclassified): is extended over $\sim$1.9~arcsec 
and contains no compact component. The object lies outside the area of
the CFHT frame. The
position given in Table~\ref{10arcmin.tab} is from the VLA-only image
(not shown).

{\bf J123734+620931} (Unclassified): is extended over $\sim$2.8~arcsec 
and contains no compact component. The object lies outside the area of
the CFHT frame. The
position given in Table~\ref{10arcmin.tab} is from the VLA-only image
(not shown).  There is a faint {\em Chandra}
counterpart.

\clearpage
\onecolumn
\section{Images of radio and sub-mm sources over optical fields}

\begin{figure}
\caption {{\bf The radio structures of the sources in the
10-arcmin field}.  The radio images are shown contoured and overlaid
on the astrometrically aligned optical fields which are displayed in
false colour (CI = --1, 1, 2, 3, 4,....10 $\times$ 10 $\mu$Jy
beam$^{-1}$). All coordinates are in J2000 and the leading 12\h ~and
+62\degr ~have been omitted from the Right Ascension and Declination
scales respectively. Where no optical field is shown, the region lies
outside either the CFHT (Barger et al.\/ 1999), or {\em HST} WFPC2
images (Williams et al.\/ 1996). The angular resolution of the radio
image (in arcsec) is indicated in Table \ref{10arcmin.tab} and in the
figure together with a hatched beam area. The origin of the optical
field shown for each source are given in Table \ref{10arcmin.tab}.}
\label{colourpix}
\end{figure}

\begin{figure}
\caption {{\bf The radio emission associated with the SCUBA 850$\mu$m
sub-mm sources in the HDF}. The combined MERLIN+VLA radio images at
the position of each SCUBA sub-mm source in the HDF and HFF (Serjeant
et al.\/ 2000) are shown contoured and overlaid on the astrometrically
aligned optical field displayed in false colour (CI = 1, 1.5, 2, 2.5,
3, 3.5, 4 $\times$ 8.5 $\mu$Jy beam$^{-1}$). All coordinates are in
J2000 and the leading 12\h ~and +62\degr ~have been omitted from the
Right Ascension and Declination scales respectively. The 3$\sigma$
SCUBA positional error circle is marked. For HDF850.1, 850.4, 850.5,
850.7, and 850.8 the optical field shown is from the {\em HST} WFPC2
(Williams et al.\/ 1996). For HDF850.2 the optical field is from the
lower angular resolution CFHT (Barger et al.\/ 1999). For HDF850.6 the
central pane is from the {\em HST} and the right centre pane is from
the {\em CFHT}. Borys et al.\/ (2003) list HDF850.4 and HDF850.5 as a
single blended source SMMJ123650+621318. The 1$\sigma$ positional
error circle for this is marked in blue over HDF850.4.}
\label{scubapix}
\end{figure}
\end{document}